\documentclass[12pt,preprint2]{aastex}
\usepackage{graphicx,emulateapj5,apjfonts}
\usepackage{natbib}
\usepackage{onecolfloat}

\newcommand{\dn}{$\rm D4000_n$}
\newcommand{\hb}{{H$\beta$}}

\newcommand{\hd}{{H$\delta_A$}}
\newcommand{\hg}{{H$\gamma_A$}}
\newcommand{\mgfep}{{$\rm [MgFe]^\prime$}}

\newcommand{\mgtwofe}{{$\rm [Mg_2Fe]$}}

\newcommand{\mgtwo}{{$\rm Mg_2$}}
\newcommand{\mgb}{{${\rm Mg}\, b$}}
\newcommand{\feone}{{$\rm Fe4531$}}
\newcommand{\fetwo}{{$\rm Fe5015$}}
\newcommand{\fethree}{{$\rm Fe5270$}}
\newcommand{\fefour}{{$\rm Fe5335$}}
\newcommand{\fefive}{{$\rm Fe5406$}}
\newcommand{\fesix}{{$\rm Fe5709$}}
\newcommand{\feseven}{{$\rm Fe5789$}}
\newcommand{\caone}{{$\rm Ca4227$}}
\newcommand{\catwo}{{$\rm Ca4455$}}
\newcommand{\gband}{{$\rm G4300$}}

\newcommand{\mtol}{\hbox{$\rm \log(M_\ast/L_z)$}}
\newcommand{\age}{\hbox{$\log(t_r/yr)$}}
\newcommand{\met}{\hbox{$\log(Z_\ast/Z_\odot)$}}
\newcommand{\sigmamtol}{\hbox{$\sigma_{\log(M_\ast/L_z)}$}}

\shortauthors{Gallazzi \& Bell}
\shorttitle{Stellar mass-to-light ratio estimates from galaxy spectra}

\slugcomment{ApJS accepted 2009 October 5}

\def\aj{AJ}%
\def\apj{ApJ}%
\def\apjl{ApJ}%
\def\apjs{ApJS}%
\def\aap{A\&A}%
\def\mnras{MNRAS}%
\def\pasp{PASP}%
\begin{document}

\def\head{

\title{Stellar mass-to-light ratios from galaxy spectra: how accurate can they be?}
\author{Anna Gallazzi$^{1}$ and Eric F.~Bell$^{1,2}$}

\begin{abstract}
Stellar masses play a crucial role in the exploration of galaxy properties and the evolution of the
galaxy population. In this paper, we explore the minimum
possible uncertainties in stellar mass-to-light (M$_\ast$/L) ratios from the assumed star formation
history (SFH) and metallicity distribution, with the goals of providing a minimum set of requirements
for observational studies.  We use a large Monte Carlo library of SFHs to study as a function of galaxy spectral type and signal-to-noise
ratio (S/N) the statistical uncertainties of M$_\ast$/L values using either absorption-line data or
broad band colors.  The accuracy of M$_\ast$/L estimates can be significantly improved
by using metal-sensitive indices in combination with age-sensitive indices, in particular for
galaxies with intermediate-age or young stellar populations. While M$_\ast$/L accuracy clearly
depends on the spectral S/N ratio, there is no significant gain in improving the S/N much above
50/pix and limiting uncertainties of $\sim 0.03$~dex are reached.  Assuming that dust is accurately
corrected or absent and that the redshift is known, color-based M$_\ast$/L estimates are only
slightly more uncertain than spectroscopic estimates (at comparable spectroscopic and photometric
quality), but are more easily affected by systematic biases.  This is the case in particular for
galaxies with bursty SFHs (high H$\delta_A$ at fixed D4000$_{\rm n}$), the M$_\ast$/L of which cannot
be constrained any better than $\sim0.15$~dex with any indicators explored here. Finally, we explore the
effects of the assumed prior distribution in SFHs and metallicity, finding them to be higher for
color-based estimates.
\end{abstract}

\begin{keywords}
{galaxies: fundamental parameters, galaxies: stellar content, methods: data analysis}
\end{keywords}
}



\twocolumn[\head]
\altaffiltext{1}{Max-Planck-Institut f\"ur Astronomie, K\"onigstuhl 17, 
D-69117 Heidelberg, Germany; \texttt{gallazzi@mpia-hd.mpg.de}}
\altaffiltext{2}{University of Michigan, 500 Church St., Ann Arbor, MI 48109, USA;
\texttt{ericbell@umich.edu}}

\section{Introduction}\label{sec:intro}
The study of the galaxy population as a function of their stellar masses has become
increasingly common and important in the last years. For a long time, galaxy
scaling relations were studied in terms of galaxy luminosities, often in the
optical regime where differences in star formation histories between different
galaxies would lead to factors of 3-10 difference in luminosity for a given stellar
mass. Comparison of galaxies with different star formation histories and similar
luminosities is therefore not a well-posed exercise, because it translates into
comparing galaxies of dramatically different stellar mass. 

This situation has significantly improved in the last decade thanks to the development and
continuous refinement of methods to estimate the stellar mass-to-light ratio ($\rm
M_\ast/L$) from galaxies spectro-photometric properties. All the various techniques involve
comparison with predictions from population synthesis models. The observational constraints
used to derive galaxy $\rm M_\ast/L$ estimates range from relatively `cheap' observations of
colors \citep[e.g.][]{BE00,papovich01,BdJ01,cole01,bell03}, 
to broad-band spectral energy distributions
(SEDs) extending from the optical or UV up to the IR
\citep[e.g.][]{borch06,walcher08,franzetti08}, to spectroscopic measurements of individual
stellar absorption features \citep[e.g.][]{kauffmann03,gallazzi05} or the full optical
galaxy spectrum \citep[e.g.][]{panter04,tojeiro07,cidfernandes07}. 

The technical development has proceeded in parallel with (and has been largely
motivated and made possible by) the rise of large homogenous spectroscopic and
photometric datasets from surveys such as the Sloan Digital Sky Survey \citep{york00},
the Two Micron All-Sky Survey \citep{2mass}, COMBO-17 survey \citep{wolf04}, VIMOS VLT
Deep Survey \citep{lefevre05}, and COSMOS \citep{scoville07}. 
The application of stellar mass estimation techniques to these datasets
has allowed the community to make rapid progress towards understanding
the properties and evolution of the galaxy population
\citep[e.g.,][]{kauffmann03b,shen03,baldry06,bell03,fontana06}.  

While it is generally acknowledged that ranking galaxies by their estimated stellar masses
rather than observed luminosities provides a more physical insight, it is also recognized
that in translating observational quantities into physical parameters estimates, we are
limited by a number of statistical and systematic uncertainties. To start with, the accuracy
of physical parameters  estimates clearly relies on the accuracy of the population synthesis
models used to interpret observations (i.e. galaxy colors or spectra). In particular
differences in the stellar libraries (either empirical or theoretical) and uncertain
(not-well observationally constrained and/or not-well theoretically understood) aspects of
particular stellar evolutionary phases can affect the predicted SED for a given star
formation history (SFH). Among these, it is recognized that thermally-pulsating asymptotic
giant branch stars (TP-AGB) can have a significant impact on $\rm M_\ast/L$ values estimated
from near-infrared (NIR) colors of young stellar populations ($\sim 0.5-2$Gyr). For
instance, $\rm M_\ast/L$ of high-redshift galaxies based on \cite{maraston05} models can be
lower by 40-60\% than those based on \cite{bc03} code because of the different flux
contribution by stars in this particular phase
\citep[e.g.,][]{maraston06,bruzual07,cimatti08}. Blue stragglers and blue horizontal branch
morphologies can instead affect (in an age-dependent fashion) the SEDs, and derived
properties, of passive galaxies. Moreover, the treatment of non-solar abundance ratios in population synthesis models can affect the
stellar ages (and hence $\rm M_\ast/L$) of early-type galaxies derived from spectral features
\citep[e.g.,][]{thomas04}.
These and other stellar populations uncertainties have been
extensively explored in \cite{conroy08,conroy09}.

Dust attenuation is a key uncertainty in stellar M/L values, especially those derived
using color information (e.g., \citealt{pozzetti07}, \citealt{fontana04}, \citealt{zibetti09} - although
dust can affect absorption features like the 4000\AA-break, \citealt{macarthur05}). It has been argued
that dust should to first order cancel out in the total M$_\ast$ estimated \citep{BdJ01}. Yet, there are a
number of caveats and possible exceptions to this simplistic picture.  In particular, for systems with
high optical depth (either globally or in patches), there can be considerable absorption of light without
large amounts of reddening \citep{deJong96,driver07,zibetti09}.  If the attenuation is patchy, moving
towards resolved mass maps, and then integrating the result to a total mass, reduces this systematic
uncertainty \citep{abraham99,zibetti09}\footnote{This is an advantage of color-based stellar M/L estimates
over spectroscopic values: the spectra to be analyzed typically are only of a part of a galaxy, whereas
imaging exists across the face of the galaxy and one can account for optically-thick parts separately, if
one wishes.}.  Modeling of the effects of dust can help \citep{deJong96,driver07},     where total thermal
infrared flux can be of some use in constraining the model \citep[e.g.,][]{popescu05,dacunha08}.

The predicted photometric properties and $\rm M_\ast/L$ of galaxies also strongly depend on the
assumed initial mass function (IMF). While it is sometimes possible to convert $\rm M_\ast/L$ values
determined assuming a given IMF into another one with a simple scaling factor when the differences
are only at masses $\la 1 M_{\sun}$, IMFs that are different at masses $\ga 1 M_{\sun}$ can alter the
galaxy SED in an age-dependent way and hence have a different global effect on the colors of
star-forming and passive galaxies \citep{conroy09}. Generally, studies of the average slope of
the high-mass ($>1$~M$_\odot$) part of the stellar IMF determine slopes consistent with Salpeter
\citep[e.g.][]{kennicutt83,BG03,hoversten08}. There are a number of studies that rely on
chemical abundances that sometimes favor slightly top-heavier IMFs
\citep[e.g.][]{worthey92,gibson97,thomas99,trager00,arrigoni09}; yet,
note that chemical modeling of galaxy evolution is uncertain (owing to uncertainties in e.g., yields,
gas infall and outflow) and such results should be viewed as being interesting but necessarily
tentative in nature.

Moreover it is generally assumed that the shape of the IMF is universal and constant in time. 
Dynamical and strong lensing estimates of total M/L on galactic scales  (i.e., $\la$ half-light
radius scales) have historically proven  compatible with a universally-applicable \citet{chabrier03}
or  \citet{kroupa01} stellar IMF \citep[e.g.,][R.\ S.\ de Jong \& E.\ F.\ Bell, in
preparation]{BdJ01,cappellari06,gallazzi06,ferreras08}; such an IMF is  motivated by local
observations of the luminosity/mass function of  star clusters, and is consistent with the statistics
of  H$\alpha$ equivalent width as a function of galaxy color for luminous galaxies
\citep{hoversten08}.  There have been a few recent arguments that, if borne out by further analysis
and observations, may suggest a time-dependent stellar IMF \citep{wilkins08,dave08,vandokkum08}.  
Furthermore, statistics of H$\alpha$ equivalent width as a  function of galaxy color/UV flux show a
tendency towards fewer very massive stars in dwarf galaxies \citep{hoversten08,meurer09}  that may
indicate a change in the galaxy-averaged stellar IMF (as opposed to the IMF of individual
clusters) in low-luminosity galaxies (see also, e.g., \citealp{weidner05}). It has to be noted
however that \cite{elmegreen06} found no evidence of the composite IMF being steeper that individual
cluster IMF, supporting the equality of the IMF from small to large scales.

Finally, we note that the derived $\rm M_\ast/L$ estimates depend on the assumed
distribution in SFHs of the models used to interpret galaxy SEDs. In particular it has
been shown that the addition of bursts of star formation on top of a continuous SFH can
produce $\rm M_\ast/L$ estimates systematically different by an amount that can vary between
$\sim$10\% and a factor of two depending on the strength and fraction of starbursts
\citep[e.g.,][]{BdJ01,drory04,pozzetti07,gallazzi08,wuyts09}. In the last few years several works
have put significant efforts in quantifying the overall statistical and systematic
uncertainties on $\rm M_\ast/L$ estimates and quantities based on them, such as stellar
mass functions and cosmic stellar mass densities \citep[to mention a
few][]{cimatti08,gallazzi08,marchesini09,LS09}. The general impression from all such works
is that it is difficult to defend $\rm M_\ast/L$ estimates to much better than 0.1~dex under
the most ideal conditions, and perhaps even worse if more extreme conditions are allowed as
discussed above (e.g. if there are systematic and/or random stellar IMF variations from
galaxy to galaxy, if there are frequent prominent bursts). 

In this paper, we take a somewhat orthogonal approach to the question of $\rm M_\ast/L$
accuracies. In particular we neglect here uncertainties related to the physics of population
synthesis models and assume that we have a `perfect' model and that the IMF is indeed
universally-applicable. We further assume that dust can be ignored, either because dealing
with dust-free systems or because it can be `perfectly' corrected for. The motivation for
making such ideal assumptions is to isolate the contribution of SFH and metallicity scatter
to the $\rm M_\ast/L$ error budget. By doing so we wish to explore in depth the following
issues: i) how does $\rm M_\ast/L$ accuracy depend on the choice of stellar population
diagnostics used to estimate its value; ii) how does $\rm M_\ast/L$ accuracy respond to
increasing spectroscopic or photometric quality; iii) how do the answers to these questions
depend on the galaxy SFH (or spectral type).
 
While at first sight our approach may seem somewhat notional, given the aforementioned
uncertainties in stellar population models and concerns about the possibility of stellar IMF
variations, yet in isolating the influence of SFH and metallicity variations on $\rm
M_\ast/L$ values one can gain insight into a number of issues. 
\begin{itemize}
\item How much effort should be placed in creating realistic and
physically-motivated grids of SFH and metallicity for use in stellar mass
estimation?  If variations in SFH and metallicity produce little scatter in 
$\rm M_\ast/L$ values, then little effort needs to be expended in producing realistic
templates.  If SFH/metallicity variations lead to large scatter in stellar M/L,
even for excellent data, then the creation of realistic priors for the SFH and
metallicity distributions acquires some urgency.
\item When should we stop trying to improve the models?  For example, if a
particular source of uncertainty in a stellar population model gives an improvement
that is much smaller than the minimum possible uncertainty from plausible
variations in $\rm M_\ast/L$ from SFH/metallicities, then one could argue that that
particular improvement in stellar population model is less urgent than other
sources of uncertainty.
\item How good does my data need to be?  An investigation of the scatter in  $\rm M_\ast/L$
from SFH/metallicity variations at a given set of colors or line indices, for a given data
quality, gives a lower limit to the $\rm M_\ast/L$ uncertainties achievable with such data. 
Given a target $\rm M_\ast/L$ uncertainty and method, one then learns what the target data
quality should be (for effects related to SFH/metallicities only).  This is less of a driver
of this work, as higher quality or complementary data may be necessary to address other
sources of uncertainty (such as stellar population synthesis model uncertainties or IMF
variations); nonetheless, this work helps to build intuition as to what kind of data quality
might be required to produce accurate $\rm M_\ast/L$ estimates, assuming that the other
sources of uncertainty are dealt with adequately.
\end{itemize}

In order to address these issues we study in some detail a large Monte Carlo library of SFHs described
in Section~\ref{sec:library} that cover the spectroscopic and photometric properties of present-day
galaxies. We randomly select a subset of models that represent different galaxy spectral types having
different SFHs, and we create mock galaxy samples by perturbing their spectra according to a given
signal-to-noise ratio (S/N).  We follow a Bayesian approach to derive the likelihood
distribution of M$_\ast$/L for these mock galaxies adopting as observational constraints several sets
of absorption indices as outlined in Section~\ref{sec:estimates}. In Section~\ref{sec:results} we
discuss the dependence of spectroscopically-derived $\rm M_\ast/L$ estimates on the absorption indices
used, on the spectroscopic S/N, on the SFH scatter, both in general and as a function of galaxy
spectral type. We then compare these results to those concerning $\rm M_\ast/L$ estimates based on
optical or optical-NIR colors in Section~\ref{sec:colors}. In Section~\ref{sec:prior} we explore
the effects on $\rm M_\ast/L$ estimates of any mismatch between the assumed SFH and metallicity prior
and the true distribution. We finally present our conclusions in Section~\ref{sec:conclusions}.
Throughout this work we adopt a \cite{chabrier03} IMF with mass cut-offs of 0.1 M$_\odot$ and 100
M$_\odot$ and we refer to the $z$-band stellar mass-to-light ratio.

\section{The method}\label{sec:method}

\subsection{The library of star formation histories}\label{sec:library}
The library of star formation histories (SFHs) that we exploit in this work is the one used in
\cite{gallazzi05} to derive stellar population parameters of SDSS galaxies. The library
consists of 150000 Monte Carlo SFHs modeled by an exponentially declining SFR (with varying
timescale and time of onset of star formation) to which random bursts of star formation can be
superposed (with varying time of onset, duration and fraction of mass produced during the burst). The
probability of having a burst is set such that 10\% of the models in the library experience a
burst in the last 2~Gyr. The metallicity varies between $0.02\times Z_\odot$ and $2.5\times
Z_\odot$ but is kept fixed along each SFH (i.e. no chemical evolution is implemented). The
galaxies in the library are also characterised by different stellar velocity dispersions for
comparison with observed absorption features whose strength depends on the velocity dispersion
broadening. A uniform prior distribution is assumed for all the parameters defining the model
galaxies. A discussion on the potential effects of the prior assumptions (in particular the
fraction of bursts) can be found in \cite{gallazzi05,gallazzi08} and will be further addressed in
Section~\ref{sec:prior} concerning M$_\ast$/L estimates.

For each SFH we measure the resulting spectral absorption features, broad-band colors,
luminosity-weighted ages and stellar mass-to-light ratios (accounting for the returned mass
fraction) by convolving \citet[hereafter BC03]{bc03} simple stellar populations (SSP) with each SFH. The models
cover the wavelength range from 91\AA~ to 160$\mu$m, with 3\AA~ resolution over the range
3200--9500\AA. 

This library purely describes the stellar continuum spectrum for each SFH, while no account is
taken of dust absorption. Stellar absorption features, used to estimate stellar mass-to-light
ratios as well as luminosity-weighted ages and metallicities, are weakly affected by dust
\citep[but see][for a detailed analysis on the dependence of absorption index strengths on
dust]{macarthur05}. We also do not include any treatment of emission lines from ionized gas in
our model library. In observed spectra of star-forming galaxies emission from ionised gas can
affect in particular the measurement of the Balmer absorption features. It is crucial that
emission lines are carefully corrected for in order to measure the true stellar absorption and
hence derive an unbiased measure of the stellar age. Medium/high spectral resolution and good
signal-to-noise ratio (S/N) are particularly crucial in this respect.

\begin{figure*}
\epsscale{2}
\plottwo{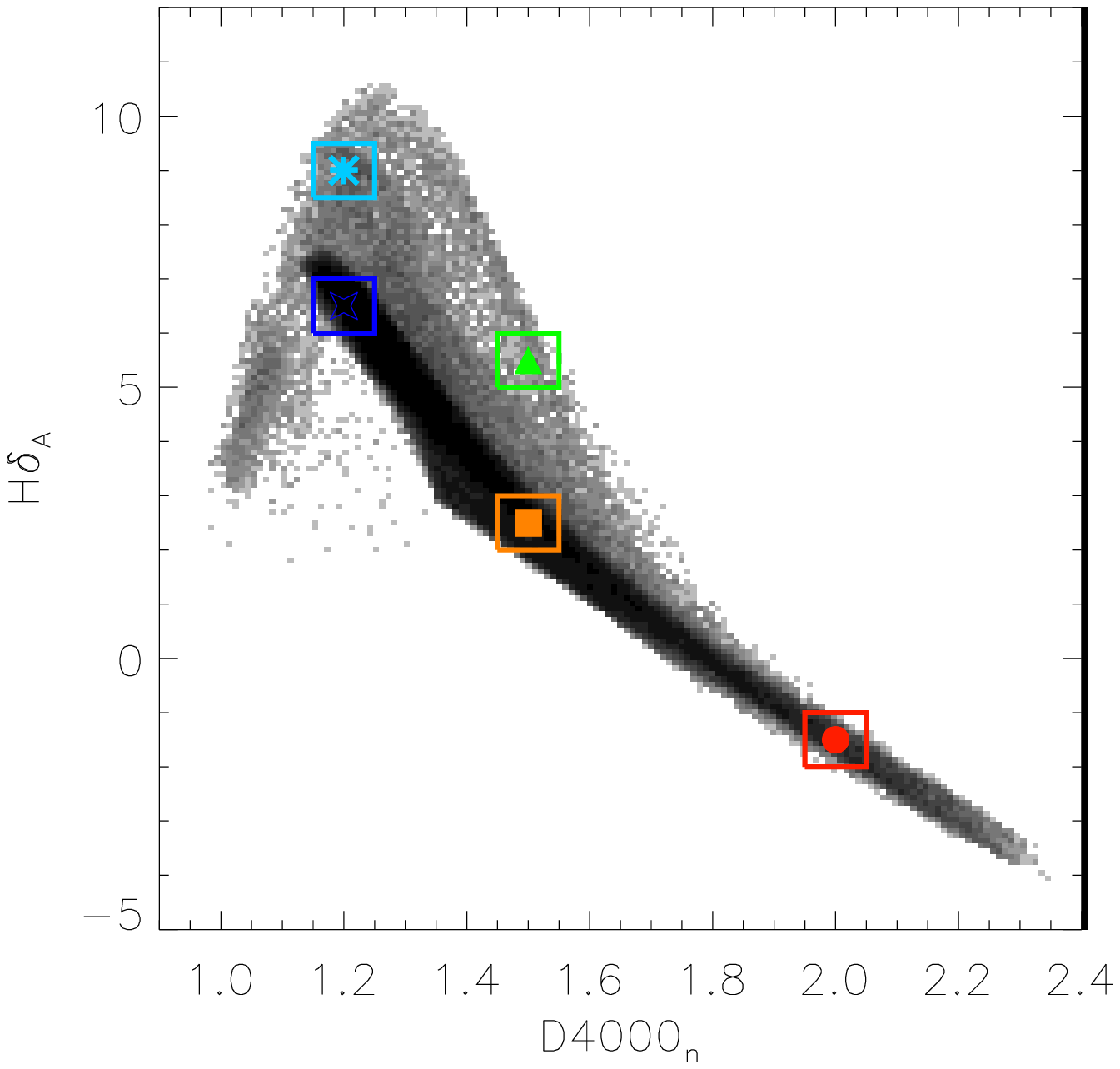}{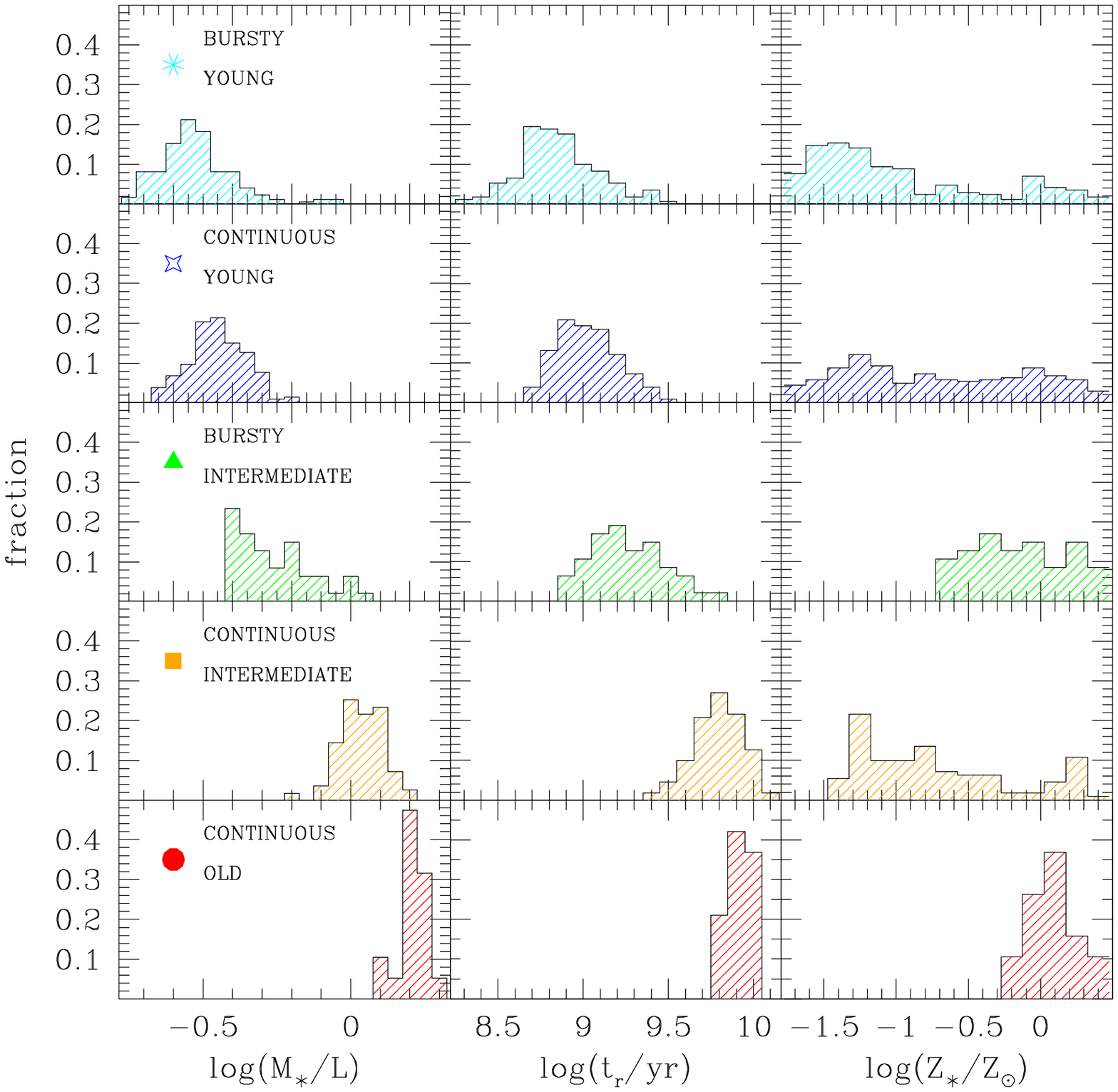}
\caption{{\it Left}: Distribution of model galaxies in the \dn-H$\delta_A$ plane. Squares
indicate the location of the analysed galaxies. Galaxies populating different positions
in the plane are characterized by different star formation histories: continuous
exponential star formation with age decreasing with decreasing \dn\ (circle: old; square:
intermediate; diamond: young) as opposed to a bursty SFH (triangle: intermediate age; asterisk:
young). {\it Right}: Stellar mass-to-light ratio (\mtol), light-weighted age (\age) and
stellar metallicity (\met) distribution of the analysed galaxies divided into the five
different SFH types.}\label{fig:distr} 
\end{figure*}

The distribution of the models in the \dn-H$\delta_A$ plane is shown in Fig.~\ref{fig:distr}
(left panel). This plane is an observational diagnostic of recent star formation history
\citep[as already discussed in, e.g.,][]{kauffmann03}. Models without bursts or which had a
burst more than $\sim6$~Gyr ago (`continuous' models, i.e. dominated by an exponentially
declining SFH) form a continuous sequence of decreasing H$\delta_A$ at increasing \dn, reflecting
an increase in the luminosity-weighted age. The region of relatively high H$\delta_A$ values at
intermediate \dn\ is occupied by models with more than one burst during their SFH (`bursty'
models) and, in particular, which experienced a burst in the last 2~Gyrs. Higher burst strengths
produce higher H$\delta_A$ values at fixed \dn, resulting in a trend of decreasing
luminosity-weighted age at increasing H$\delta_A$ at fixed \dn\ in the range $\rm 1.1\la
D4000_n\la1.8$. Finally, models that reach very low \dn\ and low H$\delta_A$ values had a very
recent ($\la100$~Myr) and intense ($\ga30$\% in mass with respect to the underlying continuous
SFH) burst of star formation. As shown in \cite{gallazzi05}, the stellar mass,
luminosity-weighted age and stellar metallicity of observed $z=0$ galaxies all increase with
increasing \dn. At intermediate \dn\ values, where the relation with stellar mass is weaker,
there is a broader range in ages, which decrease at increasing H$\delta_A$, and in metallicities,
which increase with increasing H$\delta_A$.  

In this work we wish to explore our ability of constraining $\rm M_\ast/L$ also in dependence of the
galaxy SFH (or the galaxy spectral type). For this purpose the models that we choose to analyse as
real observed galaxies are randomly drawn from five different locations on the \dn-H$\delta_A$ plane
(as indicated by the boxes in the left panel Fig.~\ref{fig:distr}): continuous models dominated by
either old, intermediate-age or young stellar populations (red, orange and blue, respectively), and
bursty models with either intermediate or young ages (green and cyan, respectively). The $z$-band
mass-to-light ratios (\mtol), luminosity-weighted ages ($\rm \log(t_r/yr)$) and stellar metallicities
($\rm \log(Z_\ast/Z_\odot)$) of the different classes of models are shown in the right panel of
Fig.~\ref{fig:distr}. There is a clear sequence of decreasing both \mtol\ and $\rm \log(t_r/yr)$ with
decreasing \dn\ and increasing H$\delta_A$ (i.e. from `continuous old' to `bursty young' types).
Besides a continuous shift in the average mass-to-light ratio and stellar age, the distributions in
these parameters tend to become slightly broader going from the homogeneous `continuous old'
populations to more complex SFHs (and more composite stellar populations). This is even more the case
when stellar metallicity is considered. While the metallicity of galaxies with strong \dn\ is
typically solar and always greater than 40\% solar (because the 4000\AA-break is somewhat
metallicity-sensitive, and high values of \dn\ can only be reached in systems with high metallicity),
the other galaxy types have a much broader metallicity distribution, even covering the whole range of
the models for the youngest populations. At intermediate ages (corresponding to intermediate \dn\
values) recent bursts of star formation produce higher metallicities than continuous models (compare
green and orange histograms), reflecting the anticorrelation between metallicity and H$\delta$ (and
age) in this \dn\ range as mentioned above.

\subsection{Stellar mass-to-light ratio estimates from spectral features}\label{sec:estimates}
As mentioned before, we create mock galaxies from a total of 553 models randomly chosen
in order to represent different spectral types and SFHs. Mock galaxies are created by assigning an
error to the absorption features typical of observed spectra with a given signal-to-noise ratio (S/N)
and by perturbing the true index strengths according to this error. This procedure is repeated 100
times for each S/N value. Each time we calculate the probability density function (PDF) of \mtol\ by
comparing the `observed' absorption indices with those predicted by all the other models in the library
with the same velocity dispersion (within $\rm 15~km~s^{-1}$). We then take the average of these 100
Monte Carlo realizations of the PDF as the final \mtol\ likelihood distribution for a given galaxy at a
given S/N.

We consider nine values of S/N (namely 2, 5, 10, 20, 30, 40, 50, 100 and 200). The typical
error of each absorption feature at a given S/N is taken to be the average error of SDSS DR4
galaxies in the redshift range $0.05<z<0.22$ that have a spectrum with the corresponding S/N
per pixel. No SDSS spectra with S/N of 100 and 200 are available. We then fit a linear
relationship between the error and the S/N (in $\log-\log$ space) and extrapolate to S/N values
of 100 and 200. We note that the average errors on the absorption features (and their
extrapolation to the highest S/N values) are almost independent of the galaxy velocity
dispersion, especially at S/N$\ga10$. The index errors estimated in this way are consistent
with those measured directly from template spectra perturbed with Gaussian noise (a reasonable
assumption at $\rm S/N\ga20$).

The PDF provides information on how plausible it is that the parameter of interest lies
within a certain range of values, given the constraints from the data (the absorption indices) on the
assumed prior distribution of models. The PDF can be characterized by its median, which we take as our
estimate of \mtol, and by a confidence interval within which \mtol\ is constrained at a certain
probability level.

Throughout the paper we quantify the uncertainty of M$_\ast$/L estimates in two ways:
i) the offset between the retrieved M$_\ast$/L value (i.e. the median of the PDF) and the true one,
which quantifies the typical bias introduced by the method, together with its rms scatter; ii) the
confidence interval given by half of the 16\%-84\% percentile range of the likelihood distribution for
each mock galaxy (which is equivalent to 1$\sigma$ range for a gaussian distribution). While the rms of
the offsets tells how much the median estimate scatters as a consequence of observational errors (so it
is in a sense an estimate of the repeatability of the measure), the 68\% range of the PDF gives the
range of acceptable values of the parameter, which depends on the sensitivity of the data to the
parameter. We note that the rms is in general lower or equal to the width of the PDF. In any case,
neither quantity can be smaller than the intrinsic scatter in M$_\ast$/L within a certain range of
observables (as we discuss in Section~\ref{sec:discret}).

We wish to explore the dependence of \mtol\ uncertainties not only on galaxy spectral type and
spectral quality, but also on the observational constraints adopted to build the likelihood
distribution. To this purpose, we consider several combinations of absorption features that
represent situations in which different portions of the optical spectrum (hence different
information on the underlying stellar populations) are exploited. The absorption features analysed
here are those included in the Lick/IDS system and defined in \cite{worthey94}, the 4000\AA-break
\dn\ as defined in \cite{balogh99} and the high-order Balmer lines \hd\ and \hg\ as defined in
\cite{wo97}. The seven sets of indices we consider are:
\begin{enumerate}
\item \dn\ and \hd\ (case 1). These are the indices used by \cite{kauffmann03}. As discussed
above, these two indices provide a diagnostic of the recent SFH. In particular, the location in
the \dn-\hd\ plane is determined in first approximation by the average stellar age, while it is
less sensitive to metallicity. They have indeed been shown to provide constraints on stellar
mass-to-light ratios and stellar ages with $1\sigma$ errors within 30\% for average SDSS
spectra, while they are not sufficient to constrain stellar metallicity
\citep{kauffmann03,gallazzi05}. The metallicity insensitivity of these indices starts to break
down for old stellar ages.

\item \dn, \hd, \hg\ and \hb\ (case 2). This set of indices is equivalent to the previous one,
but it adds information from the other Balmer lines. Even assuming that all the Balmer lines
have the same sensitivity to age (and insensitivity to metallicity) we would expect a (slight)
improvement over the use of one single line. The expected improvement could be more significant
if we consider that Balmer lines are not pure age indicators. Unlike \hb, which shows little
sensitivity to both total metallicity and abundance ratios, the higher-order Balmer lines \hd\
and \hg\ are more strongly affected by metallic lines in their bandpasses. They are thus
sensitive to element abundance ratios especially at high metallicities, in particular to
variations in abundance of C and N, respectively \citep{korn05,prochaska07,schiavon07}.  The
sensitivity of Balmer lines to metal abundances is relevant here in that it adds further
information helping in better constraining the \mtol. 

\item \dn, \hd, \hg, \hb, \mgfep\ and \mgtwofe\ (case 3). This set of indices has been used in
\cite{gallazzi05} to derive constraints on stellar metallicity, light-weighted ages and stellar
masses of SDSS DR2 galaxies. The indices \mgfep\ and \mgtwofe\ are crucial to get meaningful 
constraints on stellar metallicity. They are insensitive to element abundance ratios
\citep{thomas03,coelho07} and are
thus appropriate when comparing observed spectra of (potentially $\alpha$-enhanced) ellipticals
to scaled-solar population synthesis models. We explore here the impact of adding metallicity
constraints on the quality of \mtol\ constraints.

\item \dn, \hd, \hg, \hb, \mgtwo, \mgb, \feone, \fetwo, \fethree\ and \fefour\ (case 4). This
set of indices is equivalent to the previous one, but the individual absorption features that
define the composite \mgfep\ and \mgtwofe\ indices are used. Unlike the composite indices, the
individual Mg and Fe absorption lines are sensitive to $\alpha$/Fe abundance ratio variations.
They need to be compared to models that treat variations in individual element abundances, when
analysing real galaxies. In this case they might help narrowing further \mtol\ uncertainties
because of the additional $\alpha$/Fe information. This is not the case in this work where only
scaled-solar spectra are considered. This alternative set is used here only to explore the
effect of  additional observational constraints (but not additional astrophysical information).

\item \dn, \hd, \caone, \gband\ and \catwo\ (case 5). This set combines age-sensitive indices
and metal-sensitive indices, as in the previous two cases, but exploits only the bluest part of
the optical spectrum. It may thus be of interest for higher-redshift studies in which the red
absorption features are difficult to measure because of telluric absorption and OH skyglow. \caone\ is sensitive to C and
N as well as Ca abundance. \gband\ is particularly sensitive to C and O and only little to Fe
abundance. The only Fe-sensitive index in this group is \catwo\ which also shows an opposite
dependence on Cr and, despite its name, no sensitivity to Ca abundance \citep{TB95}. Both \gband\ and
\catwo\ are almost independent of $\rm [\alpha/Fe]$ and their strengths in observed galaxy
spectra are reasonably well fitted by the scaled-solar BC03 models \citep[see fig.~18
of][]{bc03}.   

\item \hb, \mgtwo, \mgb, \fethree, \fefour, \fefive, \fesix\ and \feseven\ (case 6). As opposed
to the previous case, with this index set we wish to explore our ability of constraining \mtol\
by using only the red portion of the optical spectrum, in particular excluding the
4000\AA-break. Only \hb\ is available as age-sensitive index. In this case, \mtol\ is
determined primarily by narrowing the metallicity range of the stellar populations. A series of
elements, including C and Fe in addition to Mg, contribute to the strength of both \mgtwo\ and
\mgb. They respond to $\rm [\alpha/Fe]$ variations, but are still reasonably well reproduced by
scaled-solar models, as opposed to $\rm Mg_1$ \citep{bc03}. \fethree, \fefour\ and \fefive\ are
all similar indicators of Fe abundance. \fesix\ shows a weaker sensitivity to Fe,
counter-balanced by its sensitivity to Ca. It is hence little affected by $\alpha$-enhancement,
which explains why it is well reproduced by BC03 models. This is also the case of \feseven\
whose strength is mostly contributed by Cr and only marginally by Fe. \citep{TB95,thomas03}.

\item \hb, \mgtwo, \fesix\ and \feseven\ (case 7). This set is equivalent to the previous one
but keeps only a limited number of indices, chosen among those that have the weakest
sensitivity to $\rm[\alpha/Fe]$ and are best reproduced by BC03 models \citep{thomas03,bc03}.

\end{enumerate}

\section{Results}\label{sec:results}
\subsection{Dependence on observational constraints}\label{sec:results1}
In this section we explore how the uncertainties on the estimated \mtol\ depend on both the
spectral quality, as expressed by the mean S/N per pixel, and the absorption features used to
constrain the likelihood distribution.

\begin{figure}
\epsscale{1}
\plotone{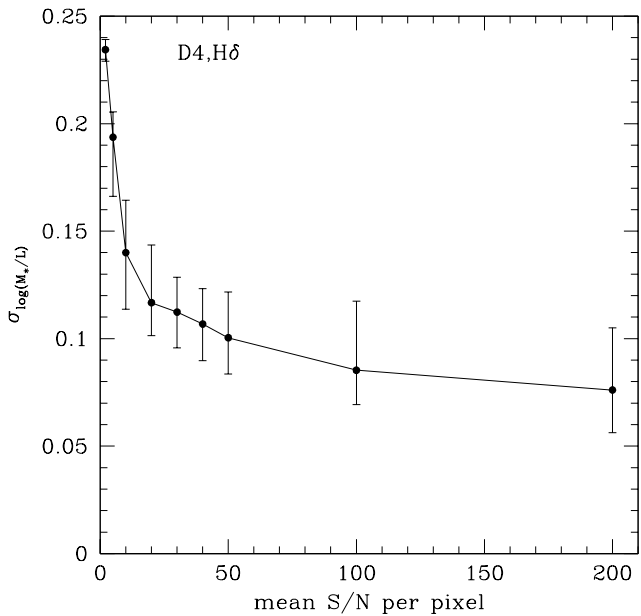}
\caption{Uncertainty on stellar mass-to-light ratio (as quantified by half of the interquartile width
of the likelihood distribution) as a function of the mean S/N per pixel of the spectrum. Points and
solid line show the median uncertainty ($\sigma_{\log(M_\ast/L)}$) of all the mock galaxies, while
error bars indicate the interquartile range of the distribution in $\sigma_{\log(M_\ast/L)}$.
Mass-to-light ratio estimates are constrained by \dn\ and H$\delta$ only.}\label{fig:case1}
\end{figure}

We first checked that for the sample as a whole the median offset between the estimated and
the true M$_\ast$/L is zero (we will discuss it for individual spectral classes in the following sections). We now
turn our attention to the median $1\sigma$ error on \mtol\ (defined as half of the $16-84$ interpercentile
range of each PDF), which is shown in Fig.~\ref{fig:case1} for all the 553 galaxies analysed as a function
of S/N. The mass-to-light ratio  estimates are obtained with only \dn\ and \hd\ (case 1). The error bars
indicate the scatter in \mtol\ error at given S/N. From the plot it is clear that the mean S/N of the
spectrum strongly influences the accuracy of spectroscopic \mtol\ estimates and a $\rm S/N\ga30$ is
required to constrain \mtol\ within better than 30\% when only two spectral indices are used. However, at
higher S/N an improvement in S/N of a factor of 2 produces only a mild improvement in the \mtol\
uncertainties. There seems to be a limiting accuracy of $\sim0.07$~dex even at S/N of 200.

\begin{figure*}
\plotone{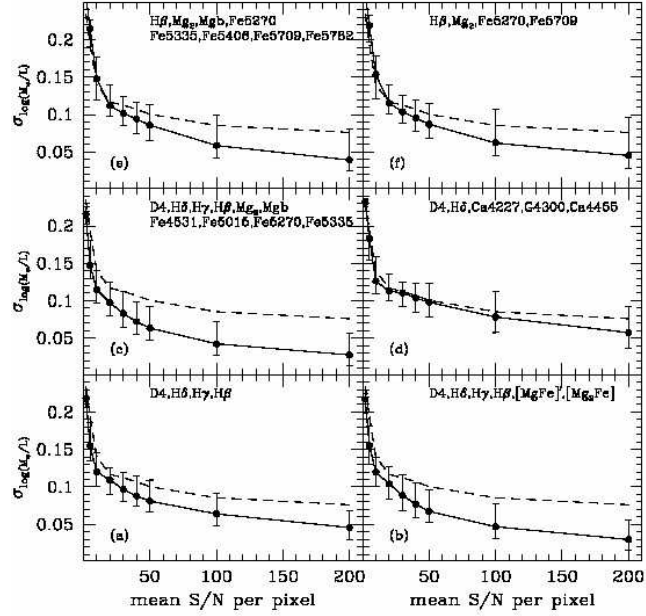}
\caption{Same as Fig.~\ref{fig:case1} but using different sets of absorption indices as
constraints, as indicated in each panel. The dashed line reproduces the curve of Fig.~\ref{fig:case1}
for the case in which only \dn\ and \hd\ are used.}\label{fig:all_cases}
\end{figure*}

In Fig.~\ref{fig:all_cases} we explore whether the uncertainty on \mtol\ estimates can be lowered
by using additional spectroscopic information. Each panel shows the $1\sigma$ error as a
function of S/N for each index set discussed in Section~\ref{sec:estimates}, in comparison to
case 1 (dashed line). Fig.~\ref{fig:all_cases}a shows that \mtol\ uncertainties can be narrowed
already by simply using all the optical Balmer lines (instead of only one) together with \dn:
the typical statistical error can go below 0.1~dex already at $\rm S/N\sim30$ and reaches 0.05~dex at the
ideal S/N of 200. 

In Fig.~\ref{fig:all_cases}b we compare the uncertainties on \mtol\ estimates derived by adding
two composite indices that are sensitive to total metallicity (case 3) with those derived in
the default case 1. The improvement is significant in particular at $\rm S/N\ga30$ and the
$1\sigma$ error can even go below the 0.05~dex level at $\rm S/N>100$. This is because, at
given age, the stellar mass-to-light ratio still depends on stellar metallicity: constraining
the range in metallicity of the stellar populations can thus help to narrow the allowed range
in \mtol. There is no difference if the individual Mg and Fe lines are used instead of the
composite $\alpha$/Fe-independent indices (Fig.~\ref{fig:all_cases}c). If observed spectra are
compared to scaled-solar models the composite indices should be used in order to avoid any bias
introduced by incorrect interpretation of $\alpha$-enhanced stellar populations. If models that
include variations in element abundance ratios are used instead, the individual indices may
give additional information not only on total metallicity
but also on relative abundances (and hence narrow the uncertainties).

In Fig.~\ref{fig:all_cases}d we explore the usefulness of other metal-sensitive lines in the blue
part of the optical spectrum, in addition to \dn\ and \hd. These features are reasonably well
reproduced by scaled-solar models, because of their weak dependence on $\rm [\alpha/Fe]$ (except
\caone). They have a negligible impact on the quality of \mtol\ estimates and there is no
improvement over the use of only \dn\ and \hd, except at extremely (and unrealistically) high S/N
ratios. Interestingly, we found similar results (better by less than a hundredth of a dex at
S/N$\ga100$) by using instead Fe lines in the same region of the spectrum, such as \feone, in
addition to \dn\ and \hd. This is due to the low resolving power of these indices and their broad
relation with $\rm M_\ast/L$. The clear improvement seen in case 3 (Fig.~\ref{fig:all_cases}b) is
mainly driven by Mg lines which alone help reducing the uncertainties by almost 0.02~dex at
S/N$\ga50$.

Finally, Fig.~\ref{fig:all_cases}e,f explore the case in which only the red portion of the optical spectrum
is available, in particular there is no measure of the 4000\AA-break. Here we use only \hb\ as
age-sensitive index and some Mg and Fe lines to constrain the metallicity range. Surprisingly these
absorption features provide \mtol\ estimates with lower uncertainties than the case in which the
reference age diagnostics (\dn\ and \hd) are used (but not better than the results obtained combining \dn\
and \hd\ with Mg and Fe lines on a longer wavelength baselines, such as case 3 or 4). The improvement,
though, becomes significant only at $\rm S/N\ga50$ and is comparable to case 2 (i.e. all Balmer lines in
addition to \dn, panel a). We note that, while metallicity constraints help to reduce \mtol\ uncertainties,
it is crucial to first determine stellar age (i.e. it is crucial to use at least an age-sensitive feature).
The comparison between panels e and f shows that only three metal-sensitive indices, in addition to \hb,
are sufficient to obtain uncertainties smaller than $\sim20$\% at $\rm S/N\ga30$ and the inclusion of other
Mg and Fe lines provides only redundant information. We note that we have chosen \fesix\ and \feseven\
because of their weak dependence on $\alpha$-enhancement that makes them appropriate for use with
scaled-solar population synthesis models. If models with variable element abundance ratios are used other
Fe lines, blueward of 5500\AA, could be used (i.e., the required wavelength range is narrower).
Interestingly \hb\ seems to perform as well as and even better than \dn\ in constraining \mtol. We checked
this also by substituting in the index set of case 7 \hb\ by \dn. Indeed the relation between \mtol\ and
\hb\ in the library is much narrower than the one with \dn. The latter is particularly broad at
intermediate \dn\ values.

Overall the \mtol\ uncertainty has a similar dependence on S/N regardless of the absorption
features used. However, at a given S/N (above at least $\rm S/N\sim10$) the statistical error
on \mtol\ estimates can be significantly improved if constraints from metal-sensitive lines on
a relatively long wavelength baseline are included. If only features blueward of $\sim5000$\AA~
are available there is no significant improvement over the use of \dn\ and \hd\ only by adding
other absorption features.

\subsection{Dependence on SFH}\label{sec:results2}
In this section we wish to explore how the uncertainties on the estimated \mtol\ depend on the galaxy
spectral type, or star formation history. In practise we wish to identify for different galaxy types
the observational conditions (i.e., spectral S/N and observational constraints) under which a minimum
accuracy can be reached. 

\begin{figure*}
\epsscale{1.5}
\plotone{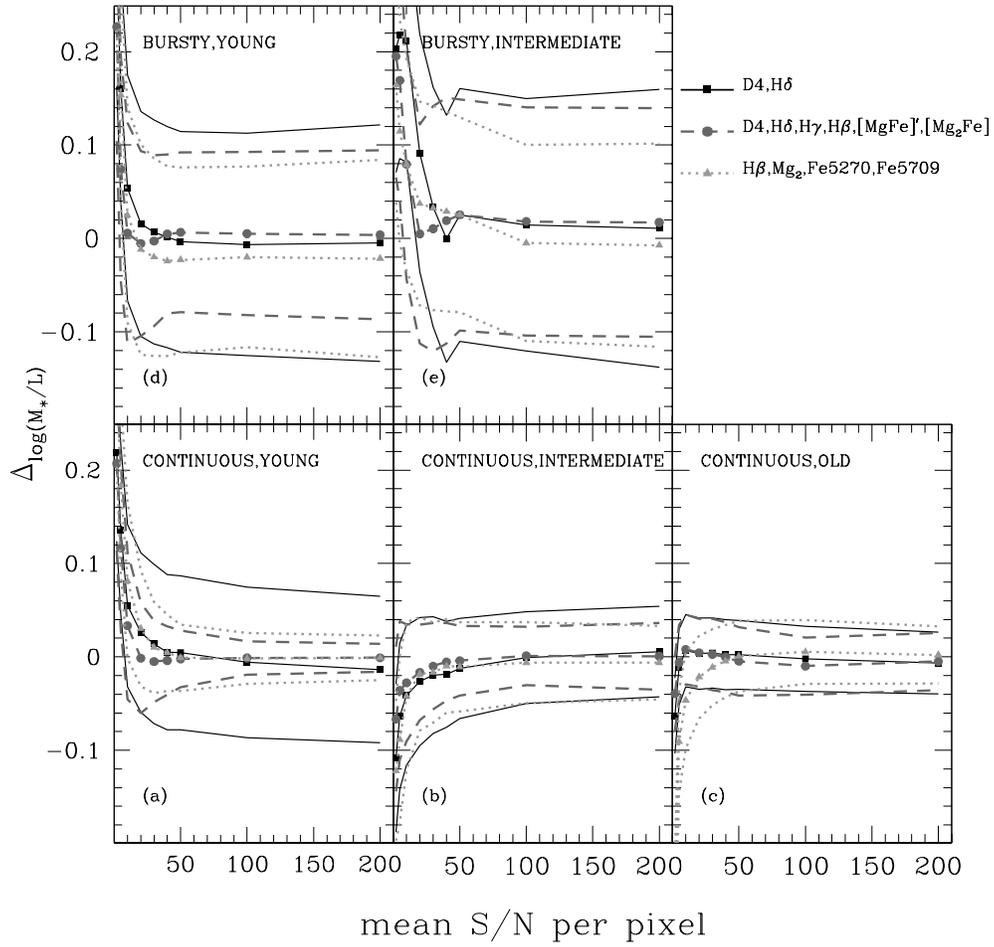}
\caption{Comparison between the estimated \mtol\ and the true \mtol\ of the model galaxy ($\rm
\Delta_{\log(M_\ast/L)}$) as a function of S/N and spectral type (as indicated in each panel). The curves
with filled circles trace the median $\rm\Delta_{\log(M_\ast/L)}$ while the lower and upper curves
represent the rms scatter of the distribution in $\rm\Delta_{\log(M_\ast/L)}$ at given S/N. As in
Fig.~\ref{fig:err_sfh}, three sets of spectral indices are shown (solid: \dn\ and H$\delta_A$; dashed:
metal-sensitive indices added; dotted: only indices in the red part of the
spectrum).}\label{fig:offset_sfh}
\end{figure*}

Before assessing the
reliability of certain observational constraints in estimating \mtol\ we need to check how well
the true \mtol\ is retrieved. Fig.~\ref{fig:offset_sfh} shows, as a function of S/N, the
difference between the estimated and the true \mtol\ ($\rm \Delta_{\log(M_\ast/L)}$) for
three combinations of absorption features and for the different
spectral types defined in Section~\ref{sec:library}: panels a, b and c represent galaxies
along the main \dn-\hd\ sequence characterized by a continuous SFH with increasing light-weighted
age (`continuous, young', `continuous, intermediate', `continuous, old'); panels d and e represent
instead galaxies with a bursty SFH (`bursty, young', `bursty, intermediate'). In each panel we
consider only three sets of indices that we deem most representative among those discussed in
Section~\ref{sec:estimates}. We consider case 1 (\dn\ and \hd, solid black line) and case 3 (\dn,
\hd, \hg, \hb, \mgfep, \mgtwofe; dashed dark grey line) because they have already been used in the
literature to derive stellar masses of large samples of galaxies and they represent two distinct
situations: case 1 uses a minimum spectral information, while case 3 seems to provide on average
the best accuracies on \mtol\ (according to Fig.~\ref{fig:all_cases}) but requires more
spectroscopic information. In addition, we consider also case 7 (dotted light grey line, using only the
red portion of the spectrum; \hb, \mgtwo, \fesix\ and \feseven): while avoiding the 4000\AA-break
region, which is most commonly used to define stellar age, it seems to bring some improvement over
the default case.

In general, all the three index combinations considered perform well in
retrieving the true \mtol\ value, at least at $\rm S/N\ga30$. At
lower S/N there can be small offsets due to the fact that models `further away' from the
analysed galaxy have a significant weight in the PDF. The index set that performs best for all
galaxy types and S/N is the one combining \dn\ and Balmer lines with metal-sensitive indices. Estimates
based on absorption features redward of the 4000\AA-break are more easily subject to biases of the order of
0.02~dex for bursty SFHs and young stellar populations.

\begin{figure} 
\epsscale{1} 
\centerline{
\plotone{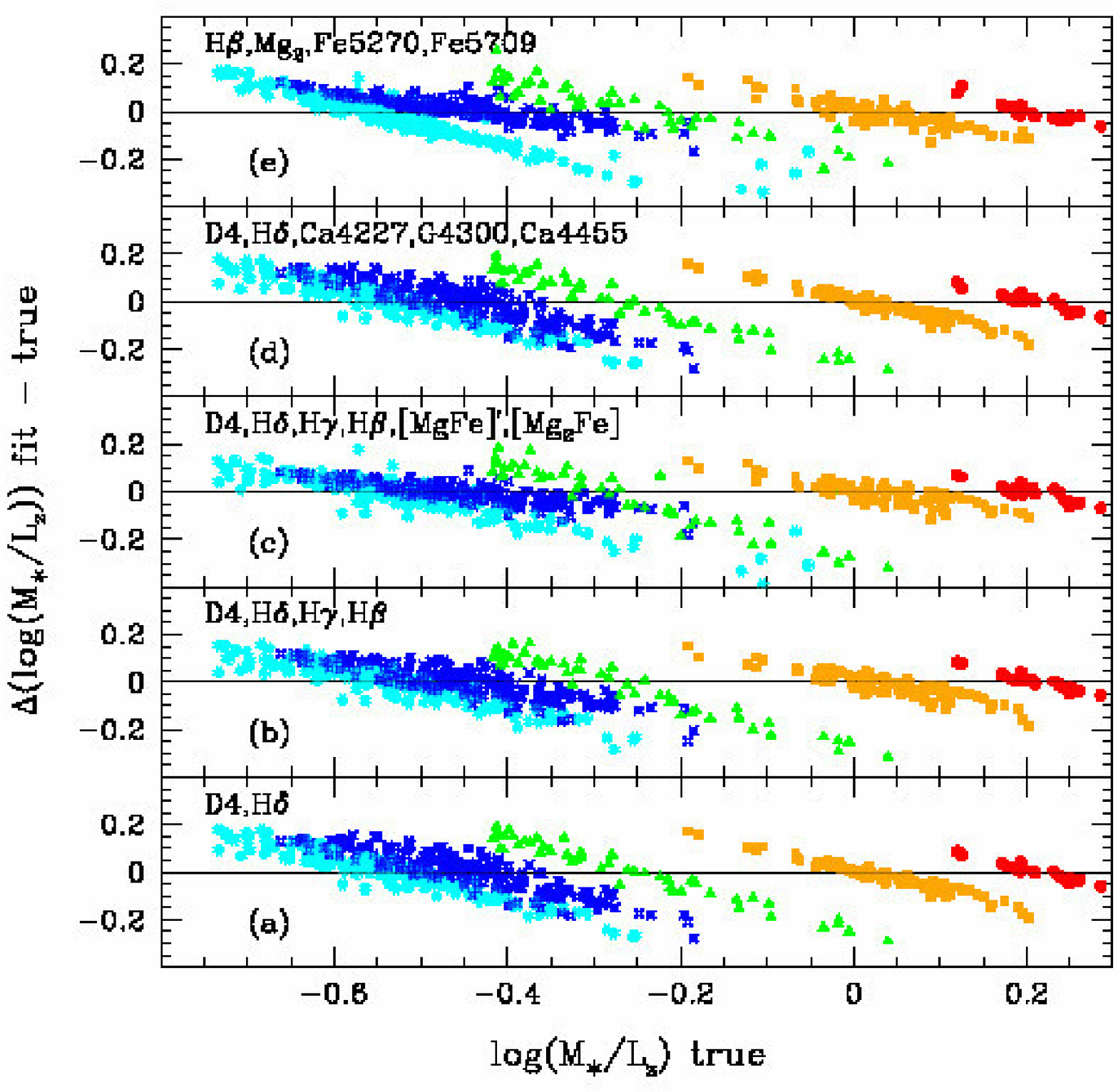} }
\caption{Comparison between \mtol\ estimates using different sets of absorption features and the
true \mtol, against the true \mtol. We do not consider here the combination of indices \dn, \hd,
\hg, \hb, \mgtwo, \mgb, \feone, \fetwo, \fethree, \fefour, and the combination \hb, \mgtwo, \mgb,
\fethree, \fefour, \fefive, \fesix, \feseven\ because they are essentially identical to case 3
(panel c) and case 7 (panel e), respectively.  Different galaxy spectral types are distinguished
using the same color coding as in Fig.~\ref{fig:distr} (red circles: `continuous, old', orange squares:
`continuous, intermediate age', blue diamonds: `continuous, young', green triangles: `bursty, intermediate
age', cyan asterisks: `bursty, young').}\label{fig:comp_idx}
\end{figure}

Fig.~\ref{fig:comp_idx} provides a further visual representation of how well the index
sets considered in this work can retrieve the true $\rm M_\ast/L$. We consider the estimates
obtained at a S/N of 30 and we distinguish the different spectral types. As already shown in
Fig.~\ref{fig:offset_sfh} there is on average no significant systematic bias in the estimated
\mtol\ values. However in many cases the offsets are correlated with the \mtol\ value itself. This
is the combined effect of the prior distribution of the models and the procedure of marginalizing
over all models. Consider the \mtol\ distribution of models with absorption features within
$1\sigma$ of the observed ones (which are those that mostly contribute to the PDF). If the analysed
galaxy has a \mtol\ value at the edges of this distribution, its \mtol\ will be over- or
under-estimated by an amount that depends on how narrow and peaked is the distribution in \mtol\ of
the models. The overall scatter and correlation between offset and \mtol\ would be reduced at
higher S/N.

\begin{figure*}
\epsscale{1.5}
\plotone{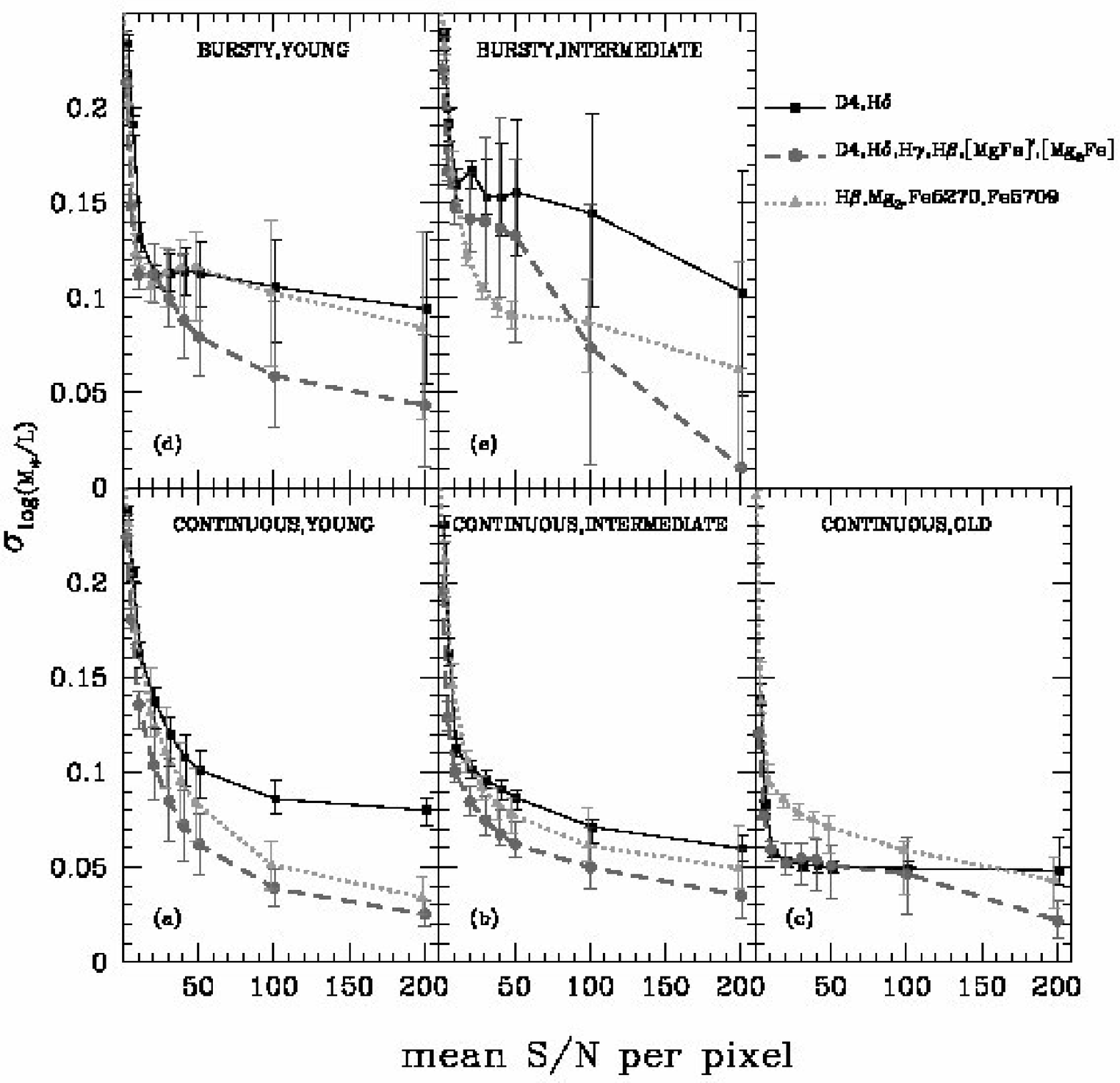}
\caption{Median uncertainties in \mtol\ as a function of S/N for galaxies populating different
regions of the \dn-H$\delta_A$ plane and characterised by different SFHs, as labeled in
each panel. Different lines styles compare the results obtained by using different sets
of absorption indices, specifically using only \dn\ and H$\delta_A$ (solid), adding
metal-sensitive indices (dashed) or using only the red part of the spectrum, i.e.
excluding the $4000\AA$-break information (dotted).}\label{fig:err_sfh}
\end{figure*}

We now analyse the statistical uncertainty on M$_\ast$/L as expressed by half of the 68\%
confidence range of the PDF (we note that this is in general larger than the rms scatter of the offsets
shown in Fig.~\ref{fig:offset_sfh} - see discussion in Section~\ref{sec:estimates}).
Fig.~\ref{fig:err_sfh} shows the median $\sigma_{\mtol}$ as a function of S/N for the five spectral types
as in Fig.~\ref{fig:offset_sfh}. It appears that the stellar M/L is generally more easily and better
constrained for galaxies dominated by old stellar populations and with smooth SFH
(Fig.~\ref{fig:err_sfh}a,b,c). This means that fewer observational constraints and poorer spectral quality
are sufficient to reach a given accuracy on \mtol\ estimates. In particular, for old stellar populations
\mtol\ can be constrained within $\pm0.05$~dex using only \dn\ and \hd\ and with a S/N of 20. Higher S/N
or further observational constraints do not improve the uncertainties, except at very high S/N by adding
metal-sensitive indices (panel c). 

Focusing on galaxies with continuous SFHs, statistical uncertainties on \mtol\ estimated with
\dn\ and \hd\ become larger, at a given S/N, with decreasing stellar population age. The
uncertainties depend much less on stellar age in case 3, i.e. when the composite
metal-sensitive indices are included. This is because the impact on $\sigma_{\mtol}$ of adding
metallicity constraints is stronger for younger galaxies, which have a much larger spread in
stellar metallicity (see Fig.~\ref{fig:distr}). Metal lines are crucial in order to reach
accuracies below 0.1~dex for galaxies with young stellar populations at reasonable S/N ratios,
and even below 0.05~dex at $\rm S/N\ga100$. It is also interesting to notice that, while \dn\
is crucial (and enough) to derive good constraints on the \mtol\ of old galaxies, for younger
stellar populations excluding \dn\ and using only absorption features redward of $\sim5000$\AA~
can even improve the uncertainty at a given S/N. As mentioned before, except at \dn$\ga2$, \mtol\
correlates more tightly with \hb\ than with \dn. 

We now turn our attention to galaxies that have experienced a recent burst of star formation,
with young ($\la1$~Gyr) or intermediate-age ($2-3$~Gyr) stellar populations (panel d and e of
Fig.~\ref{fig:err_sfh}, respectively). It is immediately clear that more observational
constraints and better spectral quality are required to reach a certain level of accuracy in
\mtol\ estimates with respect to galaxies with continuous SFHs. For `bursty young' SFHs the
only way to lower the statistical uncertainties below the 0.1~dex level is to combine age and
metallicity constraints (dashed line) and to acquire spectra with S/N above 40.

The situation is even more dramatic for galaxies in which the young burst is not dominant but
mixed to an older stellar population (panel e). The \mtol\ of this type of galaxies cannot be
constrained better than $\pm0.15-0.2$~dex. Again, a combination of \hb\ and Mg and Fe
lines performs better than including \dn\ (compare dashed and dotted lines). This class of galaxies
occupy the region where the relation between \mtol\ and \dn\ is broadest.

It is interesting to notice that, while in general higher mass-to-light ratio values are
better determined than lower ones, we have checked that, at a given S/N, the
uncertainty on \mtol\ is primarily a function of galaxy spectral type and not of \mtol\ itself.
At young ages and continuous SFHs there is a further correlation with \mtol\ in the sense that
lower \mtol\ galaxies have smaller uncertainties.

\subsection{Discreteness effects}\label{sec:discret}
At this point, we wish to spend few words to discuss the reliability of error determinations at
very high S/N. As can be seen from Fig.~\ref{fig:err_sfh}, the scatter in
$\rm \sigma_{\log(M_\ast/L)}$ at given S/N is quite small for continuous SFHs, but becomes
significantly large, and even larger than any trend of error with S/N, for bursty SFHs
especially at intermediate ages. Galaxies of similar type and with similar spectral quality can
have very different errors that may even run into the limit of zero uncertainty. Such tiny
uncertainties are not robust and are an indication that the sampling of the likelihood
distribution is not reliable. This is also suggested by rms scatter of the median
estimates (with respect to the true value) that remains significantly large even at high S/N.

All the models in the library contribute to the PDF, but the largest weights are given to those
models that lie within 1 or 2 $\sigma$ of the observational data (in this case, the absorption
indices of the analysed galaxy). It is clear then that by increasing the S/N (hence decreasing
the observational errors) and increasing the number of observational constraints (the
absorption features used to compare models and observations) fewer models enter the
`observational cell' of the analysed galaxy (i.e. have index values within $1\sigma$ of the
observed ones). The PDF is no longer uniformly sampled and the marginalization procedure
becomes effectively a maximum-likelihood analysis. The point at which this happens depends on
the number of observational data used but also on the galaxy type, or more precisely on the
location of galaxies in the index-index space. This is because of the non-uniform distribution
of the models in the \dn-\hd\ plane: while discreteness effects are not a concern for old
stellar populations and, in general, continuous SFHs (at least with such a large model library
as the one used here) even at very high S/N, they severely affect our ability of building
reliable PDFs of galaxies with bursty SFHs at $\rm S/N\ga50$ with more than two observational
constraints.

\begin{figure}
\epsscale{1}
\plotone{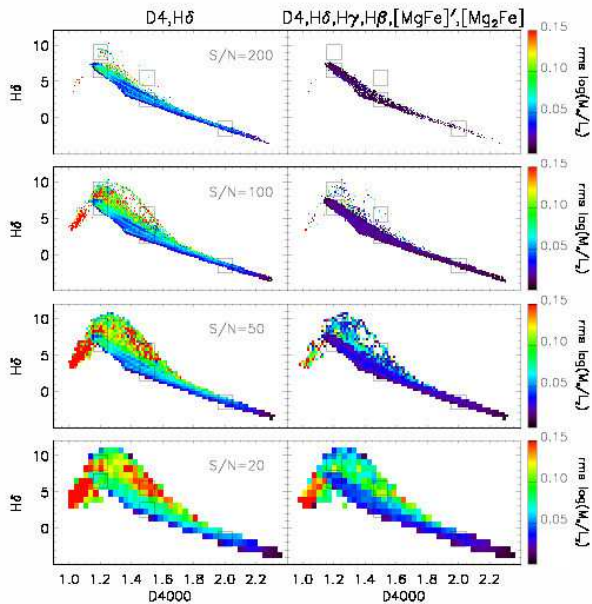}
\caption{Scatter in \mtol\ as a function of position in index-index plane and of S/N. The
left-hand panels show the \dn-\hd\ plane color-coded according to the rms scatter in \mtol\ of
models in the library in bins of \dn\ and \hd\ of width given by the observational error
typical of a given S/N. We consider here S/N of 20, 50, 100, 200 from the lower to the upper
panel. The boxes in each panel indicate the regions in which the five different spectral types
considered in this work have been defined. The right-hand panels show instead, at each \dn-\hd\
position, the average rms scatter in bins of \dn, \hd, \hg, \hb, \mgfep\ and \mgtwofe, for the
same S/N ratios as in the corresponding left-hand panel.}\label{fig:m2lscatter}
\end{figure}

Increasing the number of models (e.g. by an order of magnitude) will not improve the situation
for galaxies with high \hd\ values. The statistical error on \mtol\ cannot be lower than its
intrinsic scatter at each position in the observational space. While old stellar populations
have very uniform mass-to-light ratios, galaxies with bursty SFH have broader distributions in
\mtol. The scatter in \mtol\ remains relatively large even reducing the dimension of the
`observational cell' (i.e. reducing the index errors) and/or increasing the number of
constraints. This is illustrated in Fig.~\ref{fig:m2lscatter}, where we plot the rms scatter in
\mtol\ along the \dn-\hd\ plane for increasing S/N. We calculate the scatter in bins of \dn\
and \hd\ of width given by the typical index errors at a given S/N (left-hand panels). The
values read from this plot are in good agreement with the errors on \mtol\ shown in
Fig.~\ref{fig:err_sfh}. It is clear that the region of highest intrinsic \mtol\ scatter
(between 0.1 and 0.2~dex) is at intermediate \dn\ and high \hd\ values (except at very low \dn\
values which we do not consider in this work and are not covered by SDSS galaxies), at least up
to a $\rm S/N=100$ where there are enough models ($\ga5$) to measure the scatter.

We also calculate the \mtol\ rms scatter in bins of \dn, \hd, \hg, \hb, \mgfep\ and \mgtwofe,
and plot the average scatter in bins of \dn\ and \hd\ (right-hand panels). Although the scatter
is generally lower than in the left-hand panels (because of more observational constraints),
the variation in \mtol\ is still quite significant for bursty SFHs. 

We thus conclude that the M$_\ast$/L of intermediate-age stellar populations cannot be
determined better than $\pm0.15$~dex and there is no gain in either increasing the S/N above
$20-30$ or using more observational constraints. This is reflected in the rms of the offsets shown in
Fig.~\ref{fig:offset_sfh} that remains at the level of $0.12-0.15$~dex (depending on index set).

\section{\mtol\ estimates from colors}\label{sec:colors}
In this section we address the statistical uncertainty on \mtol\ derived from one or two
optical or optical-NIR colors. The method adopted is the same as the one described in
Section~\ref{sec:estimates} but using as observational constraints broad-band colors rather than
absorption features. It is important to remember that the model library used here does not include
dust attenuation. While this has little influence on absorption indices, observed galaxy colors are
strongly affected by dust. The uncertainties quoted here hence represent the expected
statistical uncertainties in the case of small dust attenuation (e.g. elliptical galaxies) or of
good knowledge of dust correction.

\subsection{Estimates based on one color}
\begin{figure*}
\epsscale{1.5}
\plotone{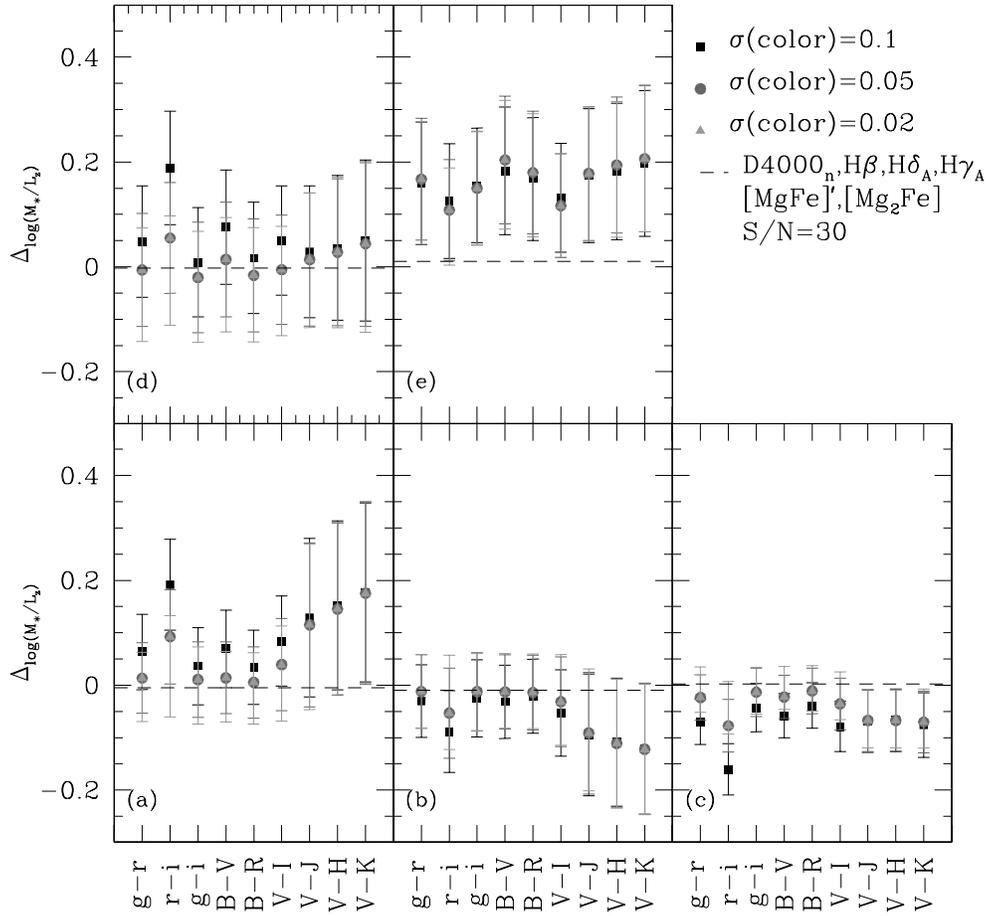}
\caption{Median offset between the estimated \mtol\ and the true value as a function of the color used as
constraint. For each color, typical errors of 0.1~mag (circles),  0.05~mag (circles) and 0.02~mag
(triangles) are considered. Each panel refers to a particular spectral type ((a): `continuous, young', (b):
`continuous, intermediate', (c): `continuous, old', (d): `bursty, young', (e): `bursty, intermediate').   
The error bars represent the rms of the distribution in $\rm\Delta_{\log(M_\ast/L_z)}$ at given S/N. The
dashed line represents the median offset for spectroscopic estimates at S/N$=30$.}\label{fig:col_offset}
\end{figure*}

We first explore the \mtol\ constraints derived using a single broad-band color. We consider in particular
the optical colors $g-r$, $r-i$, $g-i$, $B-V$ and $B-R$, and the optical-NIR colors $V-I$, $V-J$, $V-H$,
$V-K$. Fig.~\ref{fig:col_offset} shows the average difference between the retrieved \mtol\ and the
true value as a function of color, observational error and galaxy spectral type. We consider three
different values of observational error on the color, namely 0.1~mag (squares), 0.05~mag (circles) and
0.02~mag (triangles). Each panel isolate model galaxies in the five spectral classes defined in
Section~\ref{sec:library}. For comparison the dashed line in each panel traces the median offset for
estimates based on absorption indices (case 3: \dn, \hd, \hg, \hb, \mgtwofe, \mgfep) at a representative
$\rm S/N=30$ for the corresponding spectral class.

Fig.~\ref{fig:col_offset} shows that it is possible to retrieve the correct mass-to-light ratio value
within $\pm0.05$dex on average (except for bursty, intermediate-age galaxies) by using optical colors, if
their observational error is below $\la0.05$mag. They perform similarly to absorption indices in this
respect, although for old stellar populations color-based values tend to be slightly underestimated with
respect to index-based values. On the contrary, \mtol\ estimates from optical-NIR colors alone tend to
significantly under-/over-estimate the true value of old/young stellar populations. In
Section~\ref{sec:results2} and~\ref{sec:discret} we showed that spectroscopic estimates of the \mtol\ of
galaxies characterized by bursty SFH and intermediate-age populations are affected by large statistical
uncertainties and significant variation on a galaxy-by-galaxy basis. Fig.~\ref{fig:col_offset} shows that
colors systematically overestimate the true $\rm M_\ast/L_z$ of these galaxies by $0.1-0.2$dex. This is the
case for all the colors considered here.

\begin{figure*}
\epsscale{1.5}
\plotone{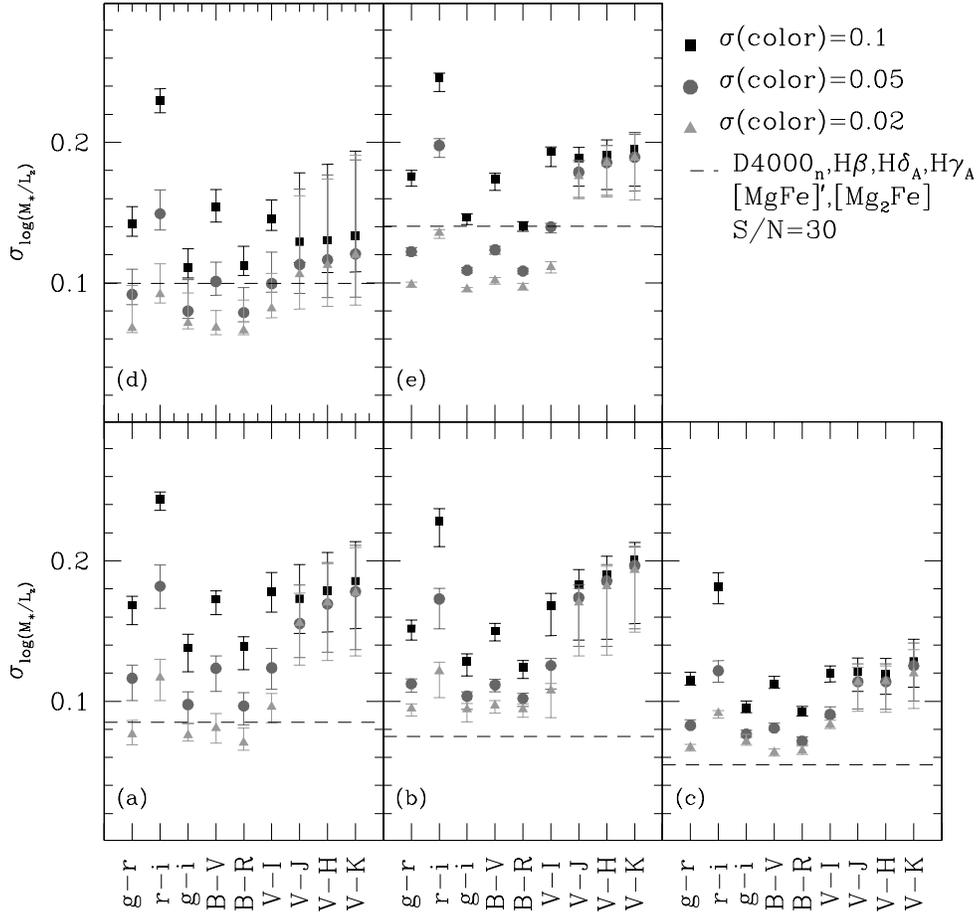}
\caption{Median error on the \mtol\ estimates derived from optical or optical-NIR colors for the five
spectral types (in the same order as in Fig.~\ref{fig:col_err}). The error bars indicate the scatter in
error ($84-16$ interpercentile range).  The dashed line in each panel indicates the typical
$\sigma_{\log(M_\ast/L_z)}$ obtained from spectral absorption features (case 3) at $\rm S/N=30$ for the
different spectral types.}\label{fig:col_err}
\end{figure*}

Fig.~\ref{fig:col_err} shows the median $1\sigma$ uncertainty on \mtol\ as a function of the color adopted.
It is interesting to see that using optical colors with reasonably small observational errors
($\la0.05$mag) it is possible to constrain $\rm M_\ast/L_z$ almost as well as using a combination of age-
and metal-sensitive absorption features (uncertainties are only $\sim0.02$dex higher), with the assumption
that redshift is known and neglecting dust corrections. Moreover the typical uncertainty \sigmamtol\ is
less dependent on the galaxy spectral class than in the case of estimates obtained with absorption indices.
For bursty SFHs the statistical uncertainty of color-based \mtol\ estimates is slightly better and
more homogeneous than those of spectroscopically-based estimates. However Fig.~\ref{fig:col_offset}
just showed that color-based estimates are systematically over-estimated by almost 0.2~dex for this type of
objects.

Fig.~\ref{fig:col_err} shows that, among the optical colors considered here, the one producing the
largest uncertainties is $r-i$ because of its shorter dynamic range and the higher relative
uncertainties with respect to the other colors. We note that observationally it is advisable, if
possible, to avoid using the $r$ band which can be severely contaminated by H$\alpha$ emission in
star-forming galaxies. On the contrary, the optical color which provides in general the smallest
uncertainties is $g-i$ which has the largest wavelength leverage. Statistical errors on \mtol\ are
below 0.1~dex for continuous models with relatively old stellar populations, and only slightly worse
for bursty, intermediate-age models ($\sim0.12$dex). This is the color preferred by \cite{zibetti09}
to derive mass-to-light ratio estimates in the $i$ band, with a maximum variation in median 
M$_\ast$/L$_i$ of $\sim0.3$dex at given $g-i$ color. 

\begin{figure}
\epsscale{1}
\plotone{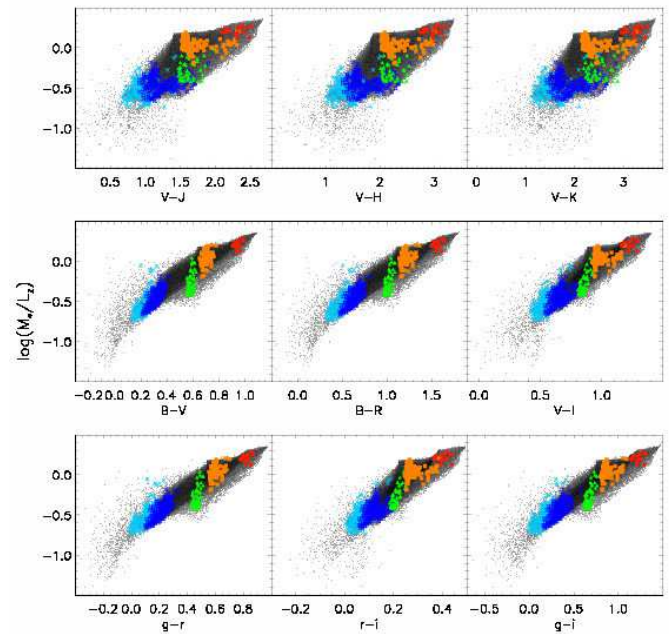}
\caption{Distribution of stellar \mtol\ as a function of color. The grey shaded region represents
the coverage of the Monte Carlo library of SFHs. Colored symbols indicate the location of the
analysed models with different \dn-\hd\ values (red circles: `continuous, old', orange squares:
`continuous, intermediate', green triangles: `bursty, intermediate', blue diamonds: `continuous,
young', cyan asterisks: `bursty, young'). }\label{fig:m2l_color}
\end{figure}

From Fig.~\ref{fig:col_err} it is also clear that optical-NIR colors predict \mtol\ with lower accuracy
than optical colors, in particular for young and intermediate-age objects (and tend to be biased as
shown in Fig.~\ref{fig:col_offset}). The relation between \mtol\ and optical-NIR colors is indeed very
broad (see Fig.~\ref{fig:m2l_color}), and optical-NIR colors are equally sensitive to age and metallicity,
as opposed to optical colors which have a stronger sensitivity to age \citep[e.g.][]{deJong96,chang06}. The
uncertainties are thus expected to be larger especially for galaxies with a young stellar population
component, which have a broader metallicity distribution (see Fig.~\ref{fig:distr}).

The trends mentioned above are clearly understood by looking at the distribution in \mtol\ as a function of
color for the whole Monte Carlo library (grey) and the analysed models (colored symbols) shown in
Fig.~\ref{fig:m2l_color}. The \mtol-optical color relation is very tight and models of different spectral
class form a sequence of increasing \mtol\ (and age) with increasing color. This and the relatively narrow
color distribution of individual spectral classes explain i) why optical colors alone can constrain \mtol\
with accuracy almost as good as the one reached from spectroscopic estimates, and ii) the small variation
in \sigmamtol\ both as a function of spectral type and within individual classes. The relation becomes
significantly broader for optical-NIR colors (which have a stronger secondary dependence on metallicity)
and the spectroscopic selection does not correspond anymore to a color selection. This explains the larger
uncertainties and the larger spread in \sigmamtol\ for estimates based on optical-NIR colors. 
Figure~\ref{fig:m2l_color} also explains the large offsets seen in Fig.~\ref{fig:col_offset}) in particular
for `bursty/intermediate' galaxies. When building the \mtol\ likelihood distribution of individual
galaxies, most of the weight comes from all the models in the library with similar color as the analysed
one and hence the derived \mtol\ will rather represent the median \mtol\ at a given color.  The offset
between the estimated and the true \mtol\ originates from the fact that `bursty, intermediate' models 
populate the low-($\rm M_\ast/L$) tail of the distribution of all models at similar color (as shown in
Fig.~\ref{fig:m2l_color}). The effect is stronger and extended to other spectral classes for optical-NIR
colors which have a broader \mtol\ distribution at fixed color.

\subsection{Comparison with Bell et al. 2003}
\begin{figure}
\plotone{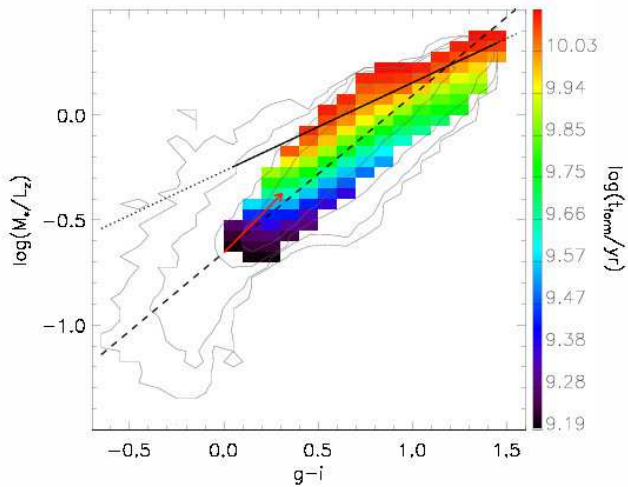}
\caption{Relation between $g-i$ color and $M_\ast/L_z$ for the model library used here. The contours
show the distribution for the library as a whole and the dashed line the relation obtained from a
simple linear fit. The solid line shows the relation given in \cite{bell03} for a 12~Gyr old stellar
population extrapolated to the full range of color covered by our library (dotted line). The
color-coded area represent the distribution of the models in our library with exponential SFH (i.e.
without bursts) as a function of the time of onset of star formation ($\rm t_{form}$). At fixed
formation time the relation between color and $M_\ast/L_z$ is shallower than the global relation
(i.e. averaged over all SFHs) and in agreement with the relation found by \cite{bell03}. The red
arrow illustrates the effect of dust attenuation on $g-i$ and the $z$-band $M_\ast/L$ assuming $\rm
A_V=1mag$ and an extinction curve $\rm A_\lambda\propto\lambda^{-0.7}$.}\label{fig:m2l_color_dust}
\end{figure}

We have also compared our Bayesian \mtol\ estimates with those obtained adopting the \cite{bell03}
formulae (see their Table 7) for SDSS colors, accounting for the different IMF. We note that, while
there is good agreement for old stellar populations, the \cite{bell03} estimates tend to be higher
than ours at bluer colors ($g-i\la1$). Indeed the \mtol-color relation of \cite{bell03} is flatter
than the one we see in our model library. This is explicitly shown in Fig.~\ref{fig:m2l_color_dust},
where we compare the \cite{bell03} relation between \mtol\ and $g-i$ (solid line; the dotted line
extrapolates their relation to the whole color range of our library) and the distribution of our
models (grey contours). The color-shaded region shows for clarity only models with exponential SFHs
(without bursts) coded according to their formation time ($\rm log(t_{form}/yr)$). As discussed in
more depth by de~Jong \& Bell (2009, in prep.), the difference in slope between our and \cite{bell03}
relation is primarily due to the age distribution of the models: at bluer colors the SFH distribution
is skewed towards substantially younger ages for the set of model templates used here compared to the
SFH distributions assumed by \cite{bell03}. The relation between \mtol\ and $g-i$ for models with
$t_{form}\ga10$Gyr is in good agreement with their relation for a 12~Gyr old stellar population. At
fixed younger formation age only the zero-point moves to bluer color, while the slope does not change.
This produces a steeper age-averaged relation (dashed line). 

Finally, we note that dust also has a secondary effect.
The red arrow in Fig.~\ref{fig:m2l_color_dust} gives an impression of the effect on $g-i$ and $z$-band
$\rm M_\ast/L$ on young stellar populations assuming $\rm A_V=1$mag and an extinction curve $\rm
A_\lambda\propto\lambda^{-0.7}$. de~Jong \& Bell (2009, in prep.) compared stellar mass-to-light ratios
calculated by \cite{gallazzi05} using these stellar population models (based on line indices) with
those derived by \cite{bell03}. While the {\it intrinsic} $\rm M_\ast/L$ (i.e. the mass-to-light ratio of
the stars alone) from both methods differ, the {\it total} $\rm M_\ast/L$ (i.e. the mass-to-light ratio
that one applies to the observed light to estimate the stellar mass, which includes the effects of
dust extinction) of both methods match well.  Dust reddens and dims the youngest stellar populations,
bringing the total $\rm M_\ast/L$ values of \cite{gallazzi05} and \cite{bell03} into better agreement. A
similar discussion of the sensitivity of the $\rm M_\ast/L$ distribution as a function of color to the
distribution in age of the library (formation time, intensity and duration of bursts) and to dust has
been recently presented by \cite{zibetti09} who use a model library similar to ours (but with a larger
fraction of recent bursts), explicitly including dust attenuation.

\subsection{Estimates based on two colors}
We expect that the uncertainties on photometric \mtol\ estimates can be reduced by combining
information from more than one color that bring additional and distinct information on the
underlying stellar populations. However, given the systematic flux uncertainties still present in
the models and the nearly-degenerate color information, it is found that using more than two colors
does not lower the uncertainties on the derived $\rm M_\ast/L$ \citep{zibetti09}. Therefore we restrict
the following analysis to the uncertainties on \mtol\ estimates based on a combination of two
colors. Specifically, we consider combinations that involve the $g-i$ color (which provides on
average the smallest uncertainties and the most stable results) in conjunction with 1) $u-g$, which
probes shorter wavelengths and brackets the 4000\AA-break, 2) $i-z$ probing redder wavelengths, or
3) $i-H$ which extends into the NIR and offers a longer wavelength leverage. We further consider
other two color combinations similar in spirit to $g-i$,$i-H$: $B-V$,$V-H$ and $V-I$,$I-K$. 

\begin{figure}
\plotone{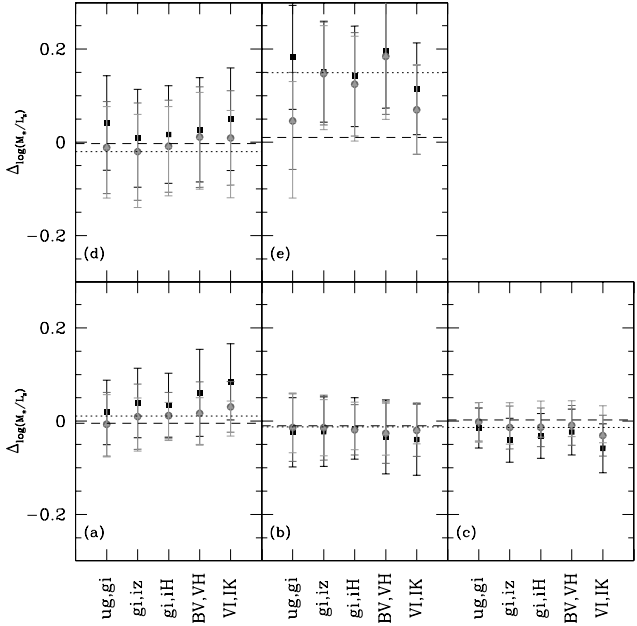}
\caption{Median offset between the estimated \mtol\ and the true value as a function of spectral type
(in the same order as in Fig.~\ref{fig:col_err}) and for different combinations of
optical/optical-NIR colors. The typical offsets for spectroscopic estimates (case 3, S/N=30) and
one-color estimates ($g-i$, 0.05~mag error) are indicated with dashed and dotted lines
respectively.}\label{fig:col2_offset}
\end{figure}

\begin{figure}
\plotone{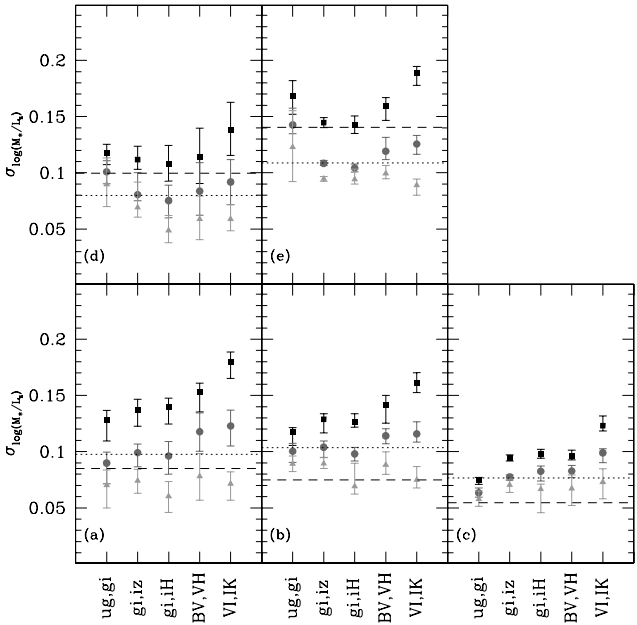}
\caption{Median error on the \mtol\ estimates derived from two colors in the optical and
near-IR. Symbols and lines have the same meaning as in
Fig.~\ref{fig:col_err}. We additionally indicate with a dotted line the median \mtol\ uncertainty
in the case of $g-i$ color with 0.05~mag error for each spectral class.}\label{fig:2col_err}
\end{figure}

Fig.~\ref{fig:col2_offset} shows the difference between the estimated \mtol\ and the true one as a
function of the five two-color combinations considered here for different spectral classes as in the
previous figures. Fig.~\ref{fig:2col_err} shows the median \sigmamtol\ and the associated scatter. We
reproduce for comparison the equivalent quantities for spectroscopic (case 3, dashed line) and one-color
estimates ($g-i$, dotted line).

From these two plots we can see that improving the photometric quality of the data have a larger
impact on the \mtol\ uncertainty associated to optical-NIR color-based estimates, probably because of
their lower degree of degeneracy, with respect to optical color-based estimates. In general we notice
that the combination of colors that performs better is $g-i$,$i-H$. However, assuming an uncertainty
of 0.05~mag on all colors, there is overall no significant improvement by using two colors with
respect to $g-i$ only (compare circles with dotted line). This is in agreement with the finding of
\cite{zibetti09} that the mass-to-light ratio in $i$-band (or equivalently in $z$-band, as used here)
has little residual dependence on optical-NIR colors. We stress that we do not include dust here,
which is expected to increase the scatter in $\rm M_\ast/L$ at fixed color, hence increase the
statistical uncertainty \citep[see][ for dust effects]{zibetti09}.

At fixed photometric accuracy, different color combinations seem to perform similarly. This is not the
case for galaxies with bursty SFHs for which optical-only colors ($u-g,g-i$) seem to provide more uncertain
estimates. In the case of bursty stellar populations, two-color estimates, even with quite poor
photometric quality, seem to be affected by statistical uncertainties similar to or smaller than
spectroscopic estimates (with reasonable S/N of 30). However, they are still affected by significant
bias.

\begin{figure}
\centerline{\plotone{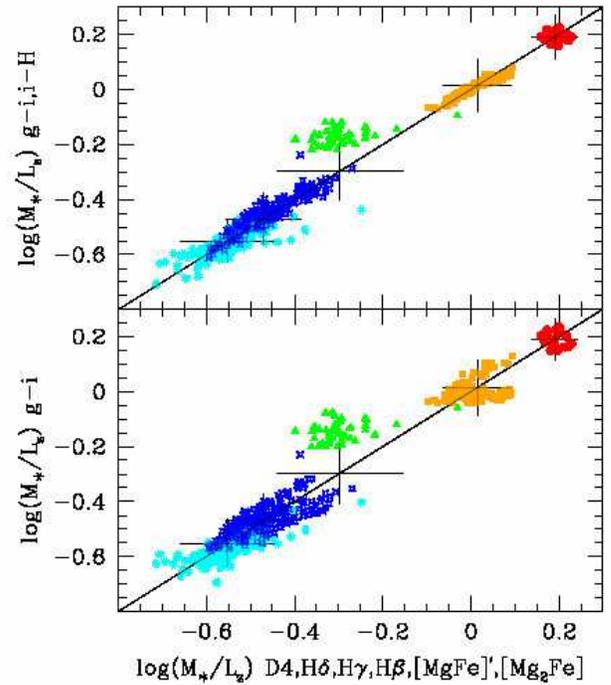}}
\caption{Comparison between \mtol\ estimated from spectral absorption features (namely \dn, \hb, $\rm
H\delta_A$, $\rm H\gamma_A$, \mgfep, \mgtwofe\ with S/N=30) and from $g-i$ color (lower panel) or
$g-i$,$i-H$ (upper panel) assuming  0.05~mag error. The solid line indicates the expected 1:1
relation. Different colors indicate different spectral types (as in Fig.~\ref{fig:distr}). The
errorbars centered at the average \mtol\ of each galaxy type indicate the average error on
spectroscopic and photometric estimates. Photometric and spectroscopic \mtol\ estimates agree well
within the respective errors, except for bursty, intermediate-age galaxy types, for which $g-i$ (as
well as other colors) tend to overpredict the true \mtol\ by $\sim0.15$~dex (see
Fig.~\ref{fig:col_offset}).}\label{fig:col_idx}
\end{figure}

Finally, in Fig.~\ref{fig:col_idx} we compare our preferred spectroscopic \mtol\ estimates, based
on a combination of age- and metal-sensitive absorption indices (case 3), with photometric
estimates from $g-i$ alone (lower panel) and the combination $g-i,i-H$ (upper panel). The plot
shows that one-color \mtol\ estimates agree well with spectroscopic ones within the respective
uncertainties. This is true for all spectral classes, except for intermediate-age models with a
bursty SFH for which we have shown that Bayesian estimates based on color tend to be biased high.
The upper panel of Fig.~\ref{fig:col_idx} shows that complementing $g-i$ with information on a
longer wavelength baseline extending to the NIR further improves the comparison with spectroscopic
estimates (and with the true \mtol) by reducing biases, although the statistical uncertainty is
not appreciably reduced. Again, we note that although the statistical uncertainty for `bursty,
intermediate' types is lower in photometric than spectroscopic estimates, adding the $i-H$ color
is not enough to remove the bias in the \mtol\ photometric estimates. This type of galaxies is
intrinsically more difficult to constrain.
 
\section{Effects of the assumed SFH and metallicity prior}\label{sec:prior}
So far we have analysed mock galaxies that are drawn from the same parent
library with which their spectra are fitted. This represents the ideal case in
which the prior distribution in SFHs and metallicities of the model library
perfectly matches the reality. Despite the very broad range of SFHs covered by
our library, this situation likely does not happen when analysing real data.
Moreover for each model SFH we assume a fixed metallicity that does not evolve
with time. This is certainly different from reality. In this section we attempt
to quantify how wrong the estimated $M_\ast$/L can be by assuming a prior that
does not match the real distribution in SFH and metallicity.

We first explore the effects on the estimated $M_\ast$/L of a mismatch in SFH
mix, in particular in the fraction of bursts of star formation. To do this we
consider two additional model libraries: i) a library of continuous models, in
which no burst is added on top of the exponentially declining SFR, and ii) a
`burst-enhanced' library, in which 50\% of the models are allowed to have a
burst in the last 2~Gyr. We then use these libraries, separately, to analyse the
mock galaxies extracted from our default library (in which 10\% of the models
have a burst in the last 2~Gyr).

\begin{figure*}
\epsscale{1.5}
\plotone{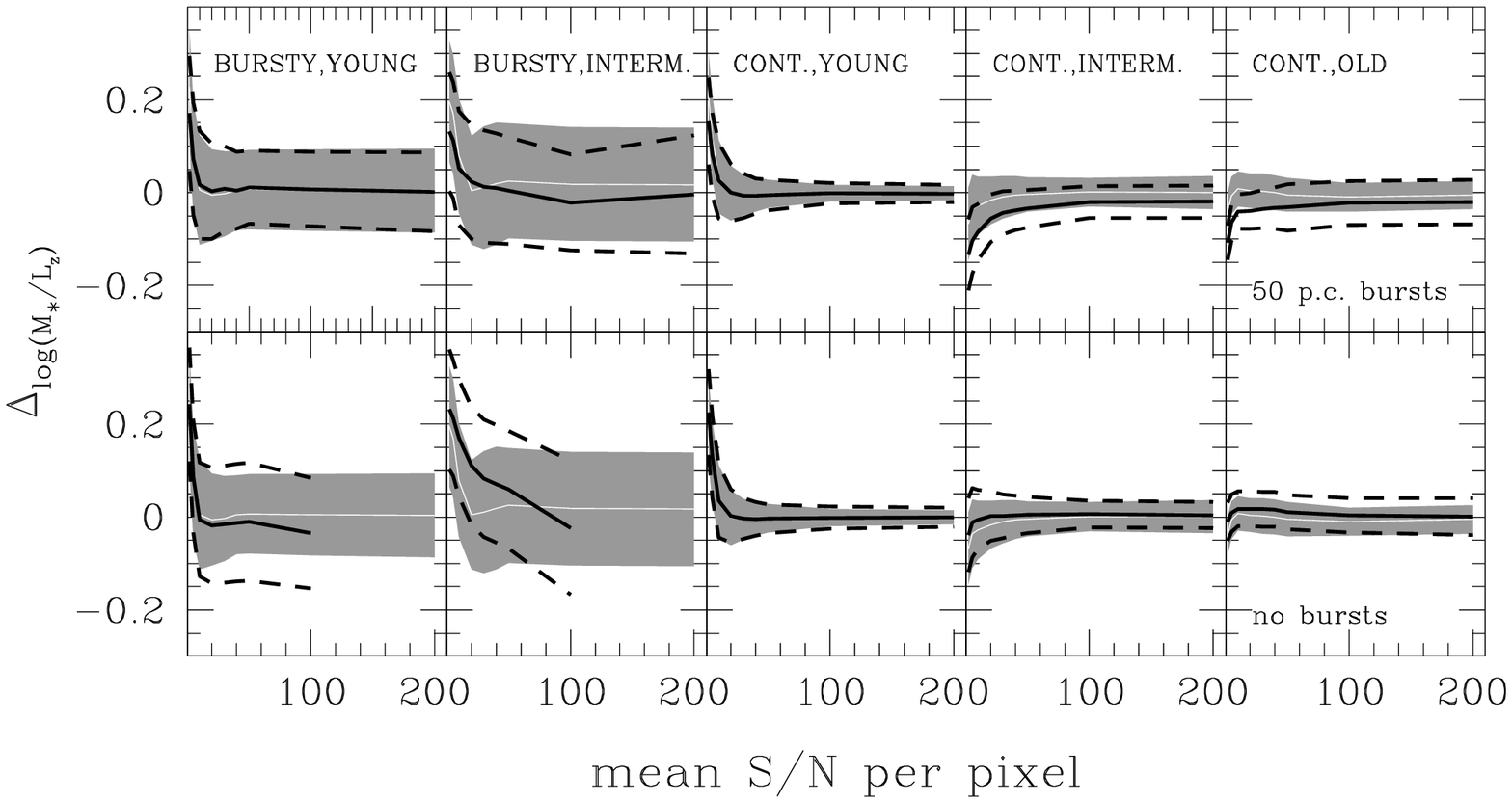}
\caption{Offset between the retrieved and the true \mtol\ as a function of S/N
for the five different spectral classes (as indicated in each panel). \mtol\
estimates are based on the spectral features \dn, the Balmer lines, \mgfep\ and
\mgtwofe. The white line and the grey shaded region indicate
the median and the rms, respectively, of $\Delta_{\mtol}$ when the mock galaxies
are fitted with the same model library from which they are drawn (as in
Fig.~\ref{fig:offset_sfh}). In the bottom panels, the solid and dashed lines
show the median and rms of the offsets for the case in which the mock galaxies
are analysed with a library without bursts. In the top panels, the black
lines represent the results of fitting the mock galaxies with a burst-enhanced
library.}\label{fig:SFH_spec}
\end{figure*}

Figure~\ref{fig:SFH_spec} shows the offset from the true \mtol\ of the value
estimated by fitting the absorption features \dn, \hd, \hg, \hb, \mgfep,
\mgtwofe (case 3) of the mock galaxies with either continuous models (left-hand
plot) or with the burst-enhanced library (right-hand plot). For comparison we
reproduce from Fig.~\ref{fig:offset_sfh} the median and rms of the offsets when
the mock galaxies are analysed with the same parent library (grey shaded
region). A library without bursts cannot reproduce high \hd\ values at fixed
\dn, i.e. the region of the \dn-\hd\ plane occupied by models with recent bursts
of star formation (see Fig.~\ref{fig:distr}). As a result, the estimated \mtol\
of `bursty' galaxies suffers of larger uncertainties and biases in particular
for intermediate-age galaxies. Clearly, the situation becomes worse by
increasing the spectral S/N. We note that in order to reproduce the observed
distribution of SDSS galaxies in the \dn-\hd\ plane it is necessary to introduce
recent bursts of star formation. As opposed to this case, a library that
over-represents recent bursts would tend to underestimate the \mtol\ of older
galaxies dominated by continuous SFH. Instead, no significant difference is seen
from the default ideal case in the median offset and rms for `bursty' galaxies.

\begin{figure*}
\epsscale{1.5}
\plotone{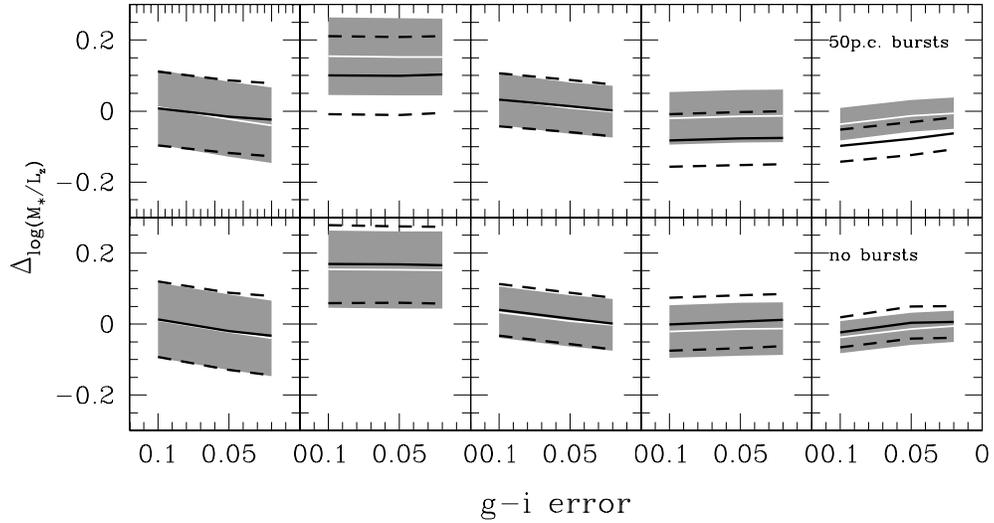}
\caption{Offset between the \mtol\ estimated by fitting $g-i$ color and the true
one as a function of the error on color (decreasing from left to right). The shaded region represents the
distribution of offsets (median and rms) in the default case as shown in
Fig.~\ref{fig:col_offset}. The solid and dashed lines show the median and
rms of the offsets when the mock galaxies are analysed with a library without
bursts (bottom panels) or with higher probability of recent bursts (top panels).}\label{fig:SFH_col}
\end{figure*}

We repeat the analysis in Fig.~\ref{fig:SFH_col} using as observational
constraint the $g-i$ color instead of the absorption features. The shaded region
represents the range of offsets in the default case (adapted from
Fig.~\ref{fig:col_offset}), while the solid and dashed lines refer to the case
in which the mock galaxies are fitted with a library without bursts (red) or
with 50\% of bursts (blue). Assuming continuous only SFHs does not introduce
additional biases in the \mtol\ estimates of `bursty' galaxies, while slightly
improves the results for `continuous' galaxies. On the other hand, an increased
fraction of models with recent bursts helps to reduce the bias in the \mtol\
estimates of `bursty/intermediate' galaxies (as can be understood from
Fig.~\ref{fig:m2l_color}), but tends to underestimate the \mtol\ of `continuous'
galaxies by $0.05-0.1$~dex.

We now explore the effects of assuming a constant metallicity along the SFH. We generate a Monte
Carlo library with the same distribution in SFHs as our default library, but allowing for chemical
evolution. The chemical evolution is modeled by a metallicity increasing from 20\% of a randomly
drawn $Z$ value (in the range $0.2-2\times Z_\odot$) until one third of the total stellar mass
produced by the underlying exponential SFH is formed, up to $2\times Z$ when two thirds of the mass
are formed. With this simple (and probably extreme) toy model of chemical evolution we do not intend
to reproduce the possibly very complex chemical evolution of real galaxies, but only to allow for a
mix in metallicities {\it within} each individual model. We note that this library does not reach
\dn\ values higher than $\sim1.9$, and hence does not properly represents observed SDSS galaxies with
higher \dn, probably because higher (mass-weighted) metallicities would be required.

\begin{figure}
\epsscale{1}
\plotone{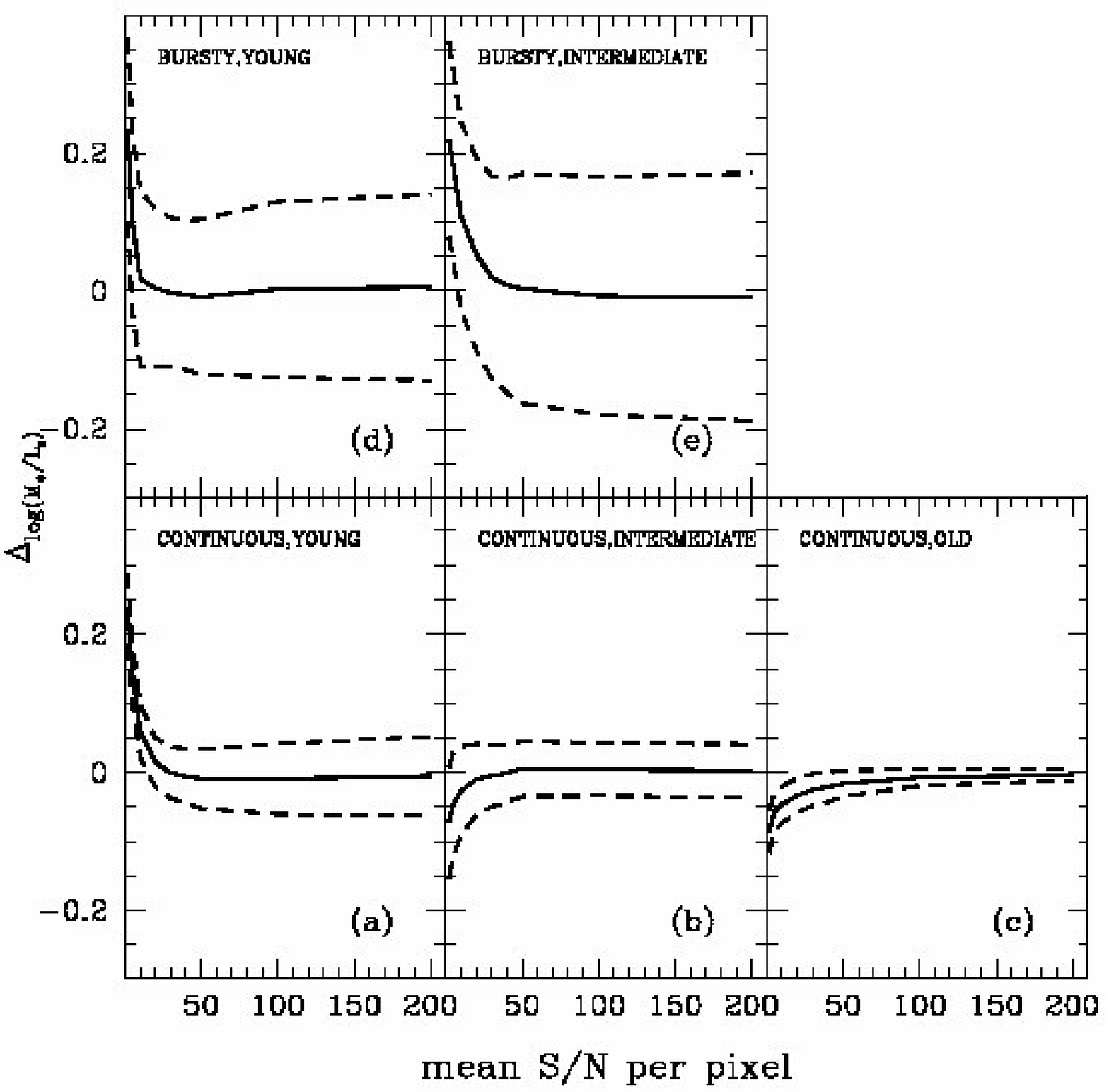}
\caption{Offset between the retrieved and the true \mtol\ as a function of S/N.
Mock galaxies are constructed allowing metallicity evolution along the SFH but
are fitted with our default library with constant metallicity. Estimates are
based on the spectral features \dn, the Balmer lines, \mgfep\ and
\mgtwofe.}\label{fig:met_spec}
\end{figure}

We now generate from this variable-metallicity library mock galaxies
representing different regions of the \dn-\hd\ plane, hence different typical
SFHs, as in the default case (we use the same ranges as in Fig.~\ref{fig:distr}
but lower \dn\ values for the `continuous/old' case). We then fit their spectral
features (case 3) with our default library in which the metallicity is constant
along the SFH. The result is shown in Fig.~\ref{fig:met_spec}, where one can see
that no significant offset in the estimated \mtol\ is introduced by assuming a
fixed metallicity, at least above S/N of $20-30$. We note however a possible
mild underestimate of the $M_\ast/L$ of old galaxies and increased uncertainties
(as shown by the rms of $\Delta_{\mtol}$) for galaxies with high \hd\ values.

\begin{figure}
\epsscale{1}
\plotone{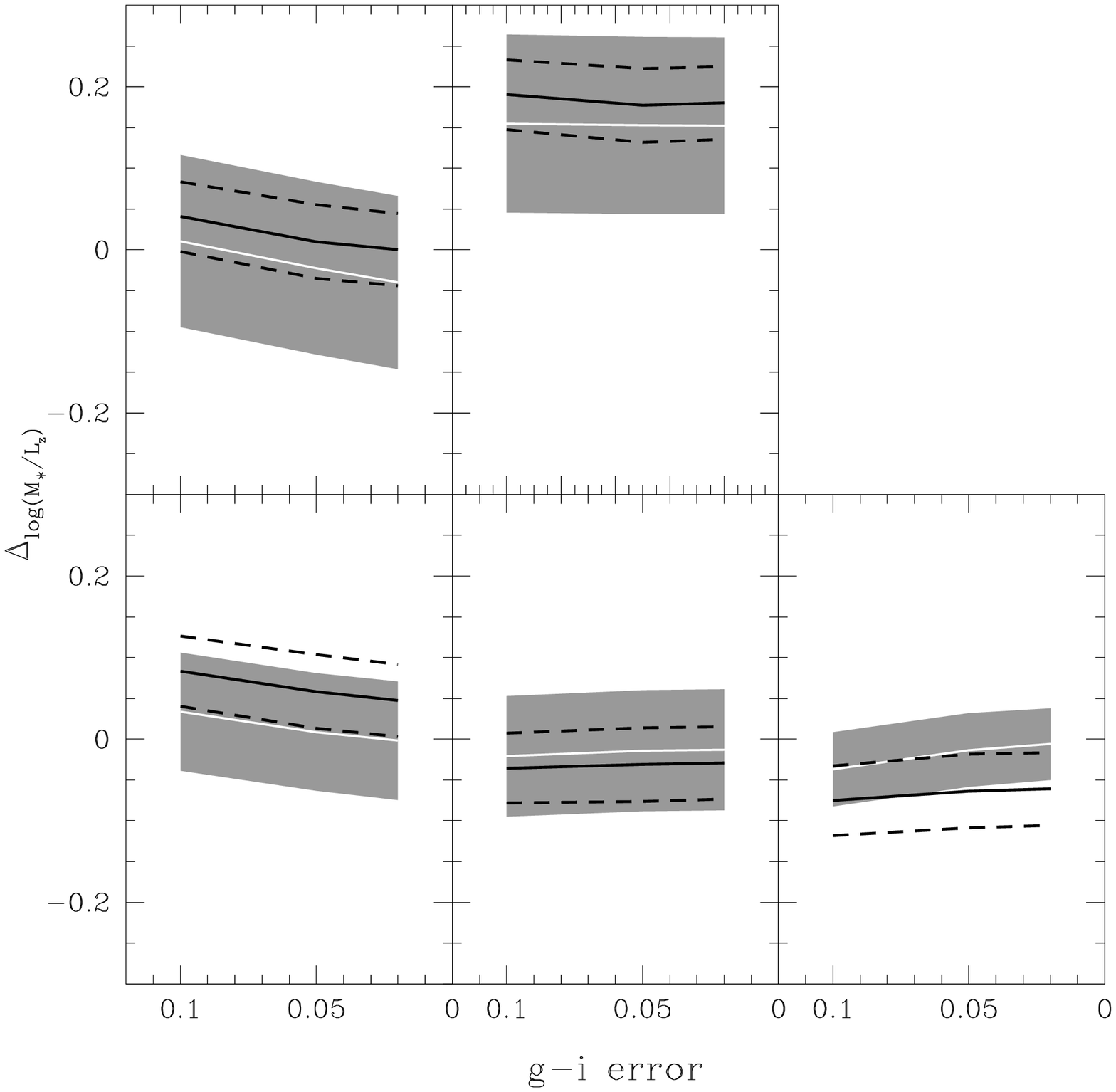}
\caption{Offset between the \mtol\ retrieved using $g-i$ color as constraint and the true \mtol\ as a
function of error on color. Mock galaxies are constructed allowing metallicity evolution along the SFH but
are fitted with our default library with constant metallicity (the solid line indicates the median
offset while the dashed lines enclose $\pm$rms scatter). The white line and shaded regions refer to the
case in which no chemical evolution is considered both in the mock galaxies and in the fitting library
(as derived from Fig.~\ref{fig:col_offset}).}\label{fig:met_col}
\end{figure}

Finally, Fig.~\ref{fig:met_col} shows the results when the $g-i$ color is used
to constrain the $M_\ast/L$ instead of the absorption features. The color of
single-metallicity stellar populations does not correctly predict the $M_\ast/L$
of composite populations with different metallicities: there is a tendency to
underestimate (overestimate) the true \mtol\ of old (young) galaxies by about
$0.05$~dex more in comparison to the default case (i.e., when fixed metallicity is
assumed both in the mock galaxies and in the fitting library).

We note that the M$_\ast$/L increases both with increasing age and with increasing
metallicity of the stellar populations. If metallicity evolves along the SFH in such a way that old
stars are metal-poorer than young ones (as in our toy model) the age and metallicity effects on
M$_\ast$/L may compensate at some level. If the age and metallicity effects on spectral indices
compensate at the same level then we should not expect to introduce a bias by assuming constant
metallicity. Indeed, we checked that M$_\ast$/L varies along the \dn-\hd\ plane, for instance, in the
same way for both the constant-metallicity and the variable-metallicity libraries, so that certain
values of absorption features are associated to the same average $M_\ast/L_z$ regardless of chemical
evolution. This is not the case for broad-band colors, which have a different response to metallicity
and age variations than M$_\ast$/L. The relation between $M_\ast/L_z$ and $g-i$ color is narrower and
slightly steeper for models with evolving metallicity than for those with fixed metallicity. This
explains qualitatively why the $M_\ast/L$ of blue (young) galaxies is overestimated, while the
$M_\ast/L$ of red (old) galaxies is underestimated.

We conclude that mismatches between the assumed and the true distribution in
SFHs can affect $M_\ast/L$ estimates: while in order to reproduce the whole range of observed
spectral features it is necessary to allow for recent bursts of star
formation, over-representing them can lead to underestimate the $M_\ast/L$ of
galaxies dominated by continuous past SFH. The effect is more significant for
color-based $M_\ast/L$ estimates. Moreover color-based $M_\ast/L$ estimates are
more susceptible of biases than spectroscopic ones if constant metallicity
is assumed.   

\section{Conclusions}\label{sec:conclusions}

\subsection{Recap}

In this work we have explored the {\it statistical} uncertainties, isolating the effects of SFH and
metallicity variations, affecting stellar mass-to-light
ratio estimates derived by comparing either stellar absorption features or colors with
predictions from a comprehensive library of SFHs. In particular we have addressed the question
of which spectroscopic constraints and which spectral accuracy (as quantified by the S/N) are
necessary/helpful to reduce the uncertainties as a function of galaxy type.

We made use of recent medium-resolution population synthesis models \citep{bc03} to generate spectral
absorption indices and broad-band colors for a Monte Carlo library of SFHs characterized by both
exponentially declining and bursty SFHs. We randomly selected $\sim550$ model galaxies in order to
cover different positions in the \dn-\hd\ plane which represent different average SFHs: continuous
SFHs following a sequence of increasing \hd\ (and decreasing stellar age) at decreasing \dn, and
bursty SFHs identified by higher \hd\ values at fixed \dn. We then generated samples of mock galaxies
at different S/N by perturbing the absorption indices of the selected models according to the typical
observational error at that S/N. For each mock galaxy we estimated the likelihood distribution of its
$z$-band mass-to-light ratio (\mtol) by comparing different sets of absorption indices with those of
all the models in the library. We quantify the accuracy on M$_\ast$/L estimates by both the
median offset from the true M$_\ast$/L and the median 68\% likelihood interval of
M$_\ast$/L computed from the PDF.

We have considered different combinations of absorption indices by including age-sensitive indices
only (\dn\ and Balmer lines), adding metal-sensitive indices (either extending to red wavelengths
in order to include Mg and Fe lines, or only in the blue portion of the spectrum $\la5000$\AA),
and using only the red portion of the optical spectrum (in particular excluding \dn).

The uncertainty on \mtol\ estimates shows a similar dependence on S/N regardless of the spectral
features used, decreasing significantly with S/N up to a S/N of 30 and then only mildly.  In
particular, for the sample as a whole, a S/N per pixel of at least $\sim30$ is required to constrain \mtol\ to within better than
30\% when using age-sensitive indices only (\dn\ and Balmer lines). However there is no significant
gain by increasing further the S/N: there seems to be a limiting uncertainty of 0.07~dex,
when using \dn\ and \hd\ only, or 0.05~dex when using all the three Balmer lines in addition to \dn.
At given S/N the best results are obtained by combining \dn\ and Balmer lines with Mg and Fe features
on a relatively long wavelength baseline, which constrain not only the age of the stellar populations
but also their metallicity (and thus narrow the allowed \mtol\ range). In this case, errors below
0.05~dex could be reached with S/N$\ga50$ spectra. When only spectra below $\sim5000$\AA~ are
available, there is no significant gain in using additional absorption features with respect to \dn\
and \hd\ only. On the other hand, when only the red portion of the spectrum is available (hence no
measure of \dn\ is possible) similar results on \mtol\ are obtained by combining \hb\ with
metal-sensitive features. 

By distinguishing galaxy spectral type, we find in general that \mtol\ is easier to constrain (i.e. lower S/N and fewer
observational constraints are required) for old stellar populations: for example, uncertainties
below $\pm0.05$dex can be reached at S/N$\sim20$ with \dn\ and \hd\ only. When using \dn\ and 
\hd\ only, \sigmamtol\ depends on stellar age in the sense that it becomes larger for galaxies with
younger luminosity-weighted ages. For such galaxies, in particular those dominated by a recent burst
of star formation, which have a broader metallicity distribution, it is crucial to add constraints
from metal-sensitive lines in order to reach accuracy $\la0.1$dex at reasonable S/N. 

The stellar mass-to-light ratio of galaxies at intermediate \dn\ and high \hd\ values,
characterized by recent bursts of SF and intermediate-age populations, cannot be constrained from
spectroscopy within better than $0.15-0.2$dex. The ability of reducing \sigmamtol\ for galaxies with
bursty SFHs (in general with high \hd\ values) is also limited by discreteness effects which come
into play at S/N$\ga50$ and using more than two observational constraints, in which cases the PDF
cannot be reliably sampled. Even if there were enough models to build a robust PDF, the intrinsic
scatter in \mtol\ at fixed index-index values is between 0.1 and 0.2~dex at intermediate \dn\ and
high \hd. We conclude that for this type of galaxies there is no gain in increasing the S/N above
$\sim30$ or using more observational constraints. 

Applying the same Bayesian method to derive \mtol\ estimates from colors shows that, in the assumption
that redshift is known and neglecting dust effects, it is possible to reach an accuracy only slightly
worse than with absorption indices. However a photometric accuracy of at least 0.05~mag is required.
The smallest uncertainties are obtained with $g-i$ which has the widest wavelength leverage. As
opposed to optical colors, optical-NIR colors of similar photometric quality provide larger
uncertainties because they show a broader relation with \mtol\ and they depend on both age and
metallicity. Moreover they tend to under-/over-estimate the true \mtol\ for old/young populations,
respectively.

We note that, while (optical) colors provide similar or smaller (and more homogeneous) uncertainties
for bursty, intermediate-age galaxies, they systematically overestimate their \mtol\ by $\sim0.2$dex.
There is no significant gain in \mtol\ accuracy by adding another color to $g-i$. Although
uncertainties are slightly smaller by combining $g-i$ with a NIR color (such as $i-H$), this is not
enough to remove the bias for bursty, intermediate-age galaxies.

\subsection{The Bottom Line}

We conclude that:
\begin{itemize}
\item The mass-to-light ratio of galaxies dominated by old stellar populations or with smooth SFH is
in general better constrained by spectroscopic absorption indices (in particular combining age- and
metal-sensitive features for young stellar populations) both in terms of retrieving the correct value
and in terms of statistical uncertainty. An accuracy below 0.1~dex is reached already at S/N$\ga20$
and it can go below 0.05~dex at higher S/N.
\item For galaxies with smooth SFH and young stellar populations  one optical color or a combination
of an optical and an optical-NIR color provides only slightly worse statistical uncertainties on \mtol\
than spectroscopic constraints, in the assumption that redshift is known and dust is not important, and
when photometric errors are below $\sim0.05$mag.
\item For galaxies which have experienced a recent burst of SF (identified by high \hd\ values at
fixed \dn) we are limited by either large uncertainties ($\sim0.15$dex) on spectroscopic estimates
or biased estimates from colors (overestimated by up to 0.2~dex).
\end{itemize}

Another important conclusion is that mismatches in the assumed SFH and metallicity distribution can
have an impact on the estimated M$_\ast$/L. It is necessary to include burst of SF in order to reproduce
the whole range of observables. However, over-representing them can lead to underestimate the M$_\ast$/L of
old stellar populations by $\la0.05$~dex if based on absorption features or up to 0.1~dex if based on
colors. Perhaps surprisingly, the assumption of constant metallicity does not affect the M$_\ast$/L
estimates from spectra, while it may introduce significant biases in color-based estimates.

\subsection{Recommendations and Practical Considerations}

We finish by highlighting a few practical issues, with some emphasis on 
responses to the particular questions posed in the introduction to this paper.

One of the prime motivations of this paper was to understand to which extent data quality affects the
determination of M$_\ast$/L. If high-quality data allowed one to uniquely select a best-fit model out
of a sample of more than a hundred thousand SFHs, then relatively little thought needs to be put into the
construction of such a model template set for M$_\ast$/Ls (in particular in the exact distribution of
different SFHs and metallicities). Unfortunately, the results of this paper have argued against this
perspective. In at least few cases, the stellar M/L values one gets depend sensitively on the exact mix of
SFHs in the model template set. The first, perhaps obvious, consideration is that the template set should
cover the full range of possible SFHs in order to be applicable to all galaxy spectral types. This becomes
critical when good quality data are available. In particular, recent bursts of SF should be well sampled in
order to analyse intermediate-age bursty galaxies, where there is considerable bias and scatter in
estimated M$_\ast$/L, even at very high S/N and using a wide range of spectroscopic indices to constrain
the model fits (summarized in, e.g., Figs.~\ref{fig:err_sfh} and~\ref{fig:col_idx}. On the other hand, the
exact fraction of bursts assumed affects the M$_\ast$/L estimates of old stellar populations
(Fig.~\ref{fig:SFH_spec} and ~\ref{fig:SFH_col}). Another case is illustrated in
Fig.~\ref{fig:m2l_color_dust}, where it is clear that the time at which star formation is assumed to start
makes a considerable difference to (at least) color-derived stellar M/L. In all these cases, any mismatch
between the mix of SFHs in the input template set and the actual Universe {\it at the redshift of interest}
will manifest itself through biases and increased scatter in stellar M/L estimates. Our analysis also shows
that, while the assumption of single-metallicity stellar populations appears not to affect spectroscopic
M$_\ast$/L estimates (as long as the range in assumed metallicities encompass the range of true
mass-weighted metallicities), it can introduce significant biases in color-based estimates. These
considerations underline one of the key conclusions of this paper, that the mix of star formation histories
and metallicities of the input template set should match as well as possible the actual distribution of
star formation histories and metallicities in order to produce optimal stellar M/L values.

A second motivation was to try to understand if and where model development acquires lower priority and
urgency, at least from the perspective of stellar mass estimates.  There are parts of parameter space
where plausible data quality could yield stellar masses accurate to $\sim 0.03$\,dex, in the absence of
any systematic mismatches between the model and data.  This potentially very high level of accuracy is
of course currently out of reach, given the systematic uncertainties inherent to current stellar
population modeling \citep[e.g.][]{maraston06,conroy08}. Accordingly, it has to be concluded that model
development and systematic error control should remain a very high priority for those wishing to
understand the stellar mass content of the Universe.

A final motivation was to understand what kind of data is necessary or sufficient  to produce
accurate stellar M/L values.   We cannot and do not offer a definitive answer to this question in
this paper, for at least two reasons.  Firstly, constraint of important sources of systematic stellar
population model uncertainty (e.g., importance of TP-AGB stars to the near-infrared light or
contribution of BHB stars to the optical light of old stellar populations) may require additional
data not considered here.  Secondly, dust has been neglected here, and further investigation may
conclude that more data will be required to model its effect properly than are required simply to
constrain the stellar populations.  Notwithstanding these caveats,  we find that, while in general
spectroscopic information is superior in constraining a galaxy SFH and hence its M$_\ast$/L, in the
case of relatively smooth star formation histories optical colors alone are enough to constrain the
stellar M/L to reasonable accuracy ($0.05-0.1$dex).  The addition of extra broad-band information
does not offer large improvements.  An obvious and important case in which the addition of spectral
information can improve the quality of M$_\ast$/L estimates is where there is highly-structured SFH
in the last $1-2$Gyr, where Balmer line information can be critical in minimizing the bias in
inferred stellar M/L value. Moreover, according to our analysis, spectroscopic M$_\ast$/L
estimates appear preferable to color-based ones in that they are less sensitive to mismatches between
the assumed and the true SFH distribution and the assumption of constant metallicity in time. 

\acknowledgements
A.G. and E.F.B wish to acknowledge support from
the Deutsche Forschungsgemeinschaft through the Emmy Noether Programme. We thank the anonymous referee for
a constructive report that helped to improve the analysis.


\end{document}